\documentclass[aps,superscriptaddress,eqsecnum,nofootinbib,showkeys,showpacs,preprintnumbers]{revtex4-2}
\usepackage{comment}
\usepackage{amsmath}
\usepackage{amssymb}
\usepackage{graphicx}
\usepackage{dcolumn}
\usepackage{bm}
\usepackage{xcolor}
  \usepackage{booktabs}

\begin{document}

\preprint{\leftline{KCL-PH-TH/2026-{\bf 12}}}

\title{Probing String-Theory–Inspired Topologies of the Early Universe through CMB Temperature and Polarization Anisotropies} 
\author{Nick E. Mavromatos}
\email{nikolaos.mavromatos@kcl.ac.uk}
 \affiliation{Physics Division, School of Applied Mathematical and Physical Sciences, National Technical University of Athens, Zografou Campus, Athens 157 80, Greece;}
\affiliation{Theoretical Particle Physics and Cosmology Group, Department of Physics, King's College London, London, WC2R 2LS, UK}

\author{Miguel-Angel Sanchis-Lozano}
 \email{Corresponding author: miguel.angel.sanchis@ific.uv.es}
\affiliation{Instituto de Física Corpuscular (IFIC), CSIC-University of Valencia, 46980 Paterna, Spain;}
\affiliation{Department of Theoretical Physics, University of Valencia, Doctor Moliner 50, 46100 Burjassot, Spain}


\begin{abstract}

TeV string-mass-scale strings have been excluded experimentally at colliders, as their effects have not been observed at the Large Hadron Collider (CERN). 
On the other hand, higher-scale string theory, with mass scales typically close to the Planck scale, is often regarded as experimentally inaccessible due to the enormous energies required for direct tests, and far beyond the reach of present or foreseeable particle accelerators. Nevertheless, the early Universe may provide an indirect observational window for high-string scale through imprints left on the Cosmic Microwave Background (CMB). In this work, building on previous studies, we reexamine temperature and polarization angular correlations as probes of the geometry and topology of the pre-inflationary Universe. We focus in particular on two-point correlation functions at large angular scales, where signatures of nontrivial spatial topology may survive as relics of primordial physics. We investigate the observational consequences of toroidal compactification and analyze their impact on the primordial power spectrum of the CMB provided by the {\it Planck} satellite. Within the current experimental and theoretical uncertainties, we identify a possible indication closely related to spatial-parity breaking, consistent with the presence of six spatial extra dimensions in the early Universe, compactified at the GUT epoch before the start of inflation. Finally, we extend our framework to B-mode polarization, highlighting its potential as a sensitive probe in forthcoming ground-based and space-borne experiments with unprecedented precision.
\end{abstract}


\maketitle

\section{Introduction}

We commence our discussion by making it clear that in this article there will be no claim of any discovery. Rather, the authors would like to report a possible hint of a signal associated with compactified extra dimensions within a string-inspired framework, inferred from the Cosmic Microwave Background (CMB) temperature and polarization anisotropies observed today. This study follows previous works \cite{Sanchis-Lozano:2025csn,sanchis-sanz:2024,Sanchis-Lozano:2022nzp,Sanchis-Lozano:2022zrp}, in which (i) imprints of Primordial Gravitational Waves (PGWs) were investigated through angular correlations, and (ii) the topology of the very early Universe was explored under the assumption of a single Kaluza–Klein extra dimension.

In conventional Kaluza–Klein or string compactifications, extra dimensions are typically associated with mass scales close to the Planck scale, rendering their direct detection inaccessible to on-Earth experiments, even at the highest energies achievable by present or foreseen particle accelerators. Nonetheless, scenarios with comparatively large extra dimensions may lead to indirect experimental signatures, for instance through missing energy and momentum in high-energy collisions at the LHC, where final-state particles could propagate into the bulk; however, no such signatures have been observed to date.


From a cosmological perspective, extra dimensions are expected to have played a dynamical role during the earliest stages of the Universe, when the characteristic energy scales approached those of fundamental unification. Although such processes occurred in the remote past, they may have left residual imprints on present-day cosmological observables. As an illustrative example, one may consider ongoing searches for signals from cosmic strings (hypothetical one-dimensional topological defects formed during symmetry-breaking phase transitions) which could have been produced either at primordial or at relatively later cosmic epochs. In the former case, strings generated shortly after inflation could induce distinctive B-mode polarization patterns in the CMB. At later times, oscillating string loops are expected to emit bursts of gravitational radiation and contribute to a stochastic gravitational-wave background, both of which are actively being sought by LIGO and NANOGrav.

In this work, we focus on the largest angular scales of the CMB temperature anisotropies. This regime remains comparatively less explored, partly because of the sizable statistical and systematic uncertainties affecting it, in contrast to the acoustic-peak region of the temperature power spectrum, where the concordance between $\Lambda$CDM theoretical predictions and observations is remarkably precise. 

Following previous analyses, we study two-point angular correlation functions of CMB photons arriving from different directions on the celestial sphere, constructed from full-sky maps. Admittedly, uncertainties at large angular scales are substantial and arise from multipole sources, including residual foreground contamination after component separation and, irreducibly, cosmic variance, which sets a fundamental limit on the precision with which these modes can be measured. Nonetheless, when the entire large-angle domain is analyzed, cumulative effects may add coherently, leading to potentially significant features whose implications will be explored in this paper.

As we shall see, several anomalies with respect to the standard cosmological model, identified in analyses of CMB data from the COBE \cite{COBE:1996}, WMAP \cite{WMAP:2013}, and {\textit Planck} \cite{Planck:2015gmu} missions, appear to be intertwined: the lack of large-angle correlations, the suppression of low multipoles in the primordial power spectrum \cite{Kim:2010st}, and an odd–even parity asymmetry \cite{Sanchis-Lozano:2022zrp}. Taken together, these features may point to a possible signature of extra dimensions, particularly within a string-motivated framework.

{\color{black} Note, however, that such parity asymmetry in the CMB, as reported by the COBE, WMAP, and \textit{Planck} missions, is distinct from, and does not contradict, the broadly assumed isotropy of the Universe embodied in the Cosmological Principle. On the other hand, its physical origin remains unclear, and the Galactic foreground contamination cannot yet be ruled out so far as a possible non-cosmological source. \cite{Land:2005jq}.} In what follows, we propose a possible explanation within the framework developed in this work.

\subsubsection{Parity breaking due to compactification of extra dimensions}

Parity imbalance between odd and even multipoles plays a fundamental role in the study of CMB angular correlations developed in this and previous work \cite{Sanchis-Lozano:2022zrp,Sanchis-Lozano:2022nzp}. It is therefore useful to first review the various mechanisms through which symmetries can be broken in the context of elementary particle physics.

In Quantum Field Theory, three main mechanisms of symmetry breaking are well known \cite{Donoghue:2022wrw}: (i) {\it spontaneous symmetry breaking}, where the Lagrangian is symmetric but the vacuum is not (e.g., the Higgs mechanism); (ii) { \it explicit symmetry breaking}, where symmetry-violating terms appear directly in the Lagrangian (e.g., a small mass term in a classically chiral-invariant theory); and (iii) {\it anomalies}, where a classical symmetry is broken in the quantization process, as in the axial anomaly responsible for the decay $\pi^0 \to \gamma \gamma$.

On the other hand, an additional mechanism of symmetry breaking, geometric in origin, can arise in higher-dimensional field theories through compactification and/or boundary conditions (BCs) in the extra dimensions. This is known as the \emph{Scherk-Schwarz} mechanism~\cite{Scherk:1979zr}, and it has been widely applied in the context of supersymmetric theories~\cite{Barbieri:2001dm}. Our approach will follow the main lines of this fourth mechanism of symmetry breaking within a string-inspired scenario.

 It is worth noting that a close analogue in condensed matter appears in topological phases, where geometry and boundary conditions play a role similar to Scherk–Schwarz twists \cite{Hasan:2010xy}. In this context, global properties of the system, encoded in its topology and boundary conditions, effectively modify the electronic band structure, leading to shifts in energy levels and the possible opening of gaps without altering the local Hamiltonian. This provides a geometric mechanism akin to symmetry breaking in higher-dimensional theories.

\subsubsection{Geometry and topology of extra dimensions and the rise of Infrared cutoffs}

Starting from a primordial scalar field in the early Universe, we adopt essentially the same assumptions as in \cite{sanchis-sanz:2024,Sanchis-Lozano:2025csn}, whereby infrared (IR) cutoffs arise naturally from the geometry and topology of the early Universe. {\color{black} In Appendix \ref{sec:appA}, we provide a general physical motivation for considering such cutoffs, which suppress modes below a characteristic scale essentially determined by the inverse of the radius of the Kaluza–Klein (KK) extra dimension.}

The key difference here is that, instead of a single characteristic energy (or inverse-length) scale, we consider multiple extra dimensions, each associated with its own characteristic scale determined by the corresponding compactification radius. In particular, we assume a toroidal compactification and impose BCs along the different compact dimensions. This leads to distinct IR cutoffs for odd and even modes, which separately affect the corresponding odd and even multipoles of the primordial power spectrum, with fundamental consequences for the angular correlations of the CMB.

 Hereafter, $N_{\rm tori}$ will denote the number of tori, each torus, of course, corresponding to a pair of extra dimensions, and $N_{\rm extra}$ the total number of compactified extra dimensions, with
 \begin{align}\label{NtoriNext}
 N_{\rm extra} = 2N_{\rm tori}.
 \end{align}
Specifically, in our approach, developed in Section \ref{sec:torcomp}, we shall deal with a (super)string toroidal compactification model to (3+1)-dimensions with $N_{\rm tori}=3$, . 



\subsubsection{Layout of the Article}

The structure of this article is as follows. In Section \ref{sec:torcomp}, we review the basic features of toroidal-orbifold compactifications of ten-dimensional superstrings down to (3+1) large (uncompactified) spacetime dimensions, and show how the aforementioned infrared cutoffs arise in the spectrum of a scalar field (\emph{e.g.}, the inflaton in realistic string-inspired cosmological scenarios).  

In Section \ref{sec:angspec}, we revisit a previous work, which was limited to the assumption of a single KK extra dimension. We discuss the introduction of IR cutoffs in the angular power spectrum and their impact on the closely related CMB temperature correlations. In particular, we analyze the contributions of scalar and tensor perturbations to the two-point angular auto-correlation function, emphasizing their differences. In Section \ref{sec:Emode}, we study the E-mode polarization of the CMB as an independent probe of temperature correlations.

Sections \ref{sec:newanal} and \ref{sec:Nextra} contain the core of the present work, where a new analysis of the large-angle anisotropies is performed within the theoretical framework developed in Section \ref{sec:torcomp}. As a suggestive result, we find that a number of extra dimensions equal to 6 is favored by our analysis, in good agreement with expectations from string theory.
 
In Section \ref{sec:bmode}, we discuss prospects for incorporating B-mode polarization into our IR cutoff analysis, which, once detected, will provide crucial evidence for primordial gravitational waves. Finally, Section \ref{sec:concl} contains our conclusions stressing the relevance of parity breaking in the CMB observations. 

Some notational aspects of our approach, particularly those related to orbifold compactification, are presented in Appendix \ref{sec:appA}, together with a short description of the code used for our numerical computations, for the convenience of the reader. On the other hand, the impact of the smoothing procedure of the infrared cutoffs on odd and even multipoles is presented in Appendix \ref{sec:appB}.

\section{Toroidal compactification in String theory}\label{sec:torcomp}

\begin{figure}[t]

\centering

\includegraphics[width=0.8\textwidth]{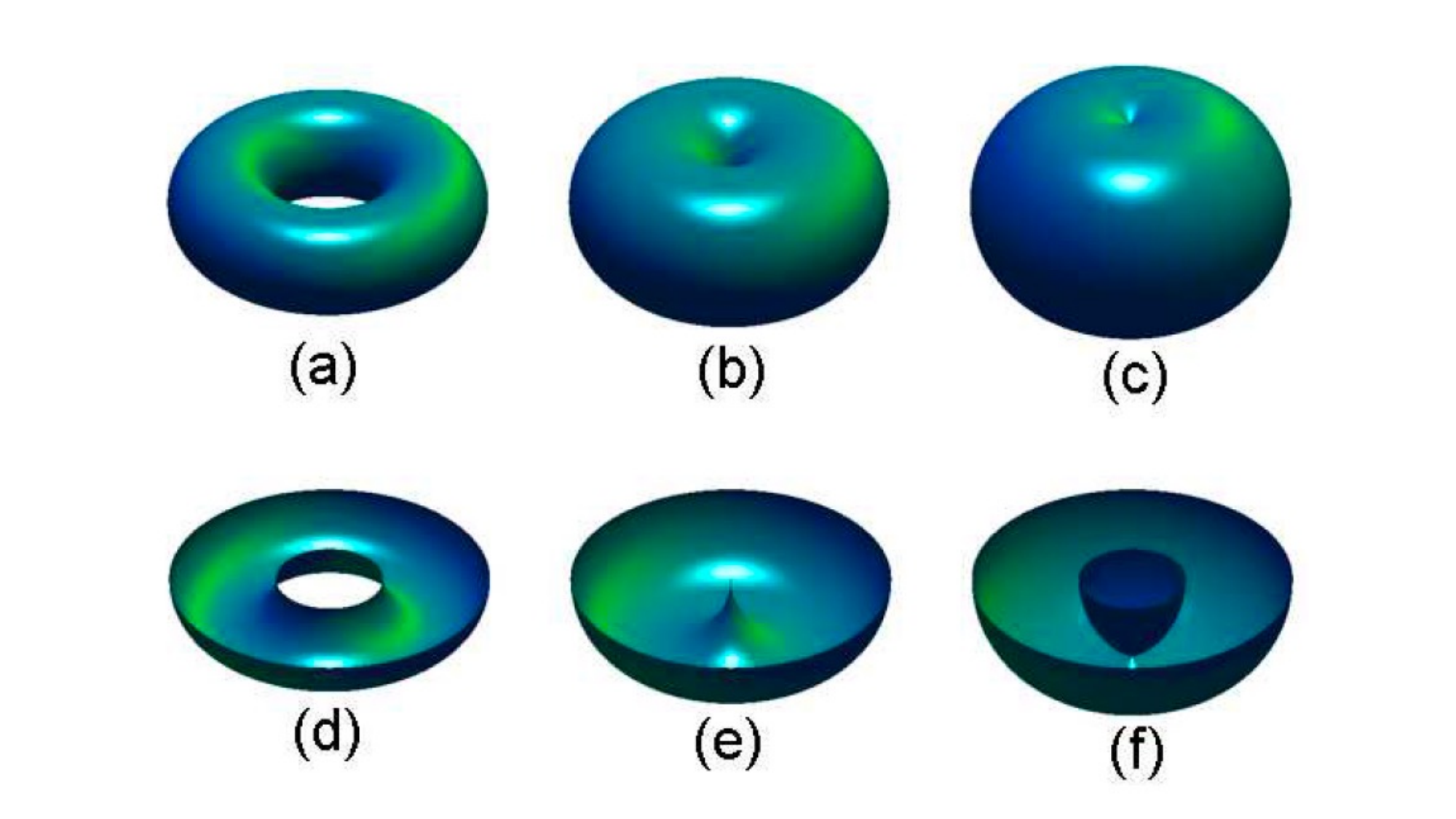}
\caption{\label{fig:tori} \underline{Upper panel}: Various categories of two-dimensional tori, classified by the value of the parameter $\gamma$, defined in \eqref{ratioR}. From left to right: (a) ordinary doughnut-shaped torus ($\gamma > 1)$, (b) horn torus without a hole ($\gamma=1$), and (c) spindle torus ($0 < \gamma < 1$, a self intersecting torus.
 \underline{Lower panel} The  cutaways, corresponding to the manifolds depicted in the upper-panel, which 
 divide the tori in half. The names of the tori are derived from the shape of the  corresponding cutaways.
 From left to right: Cutaway of a: (c) ordinary doughnut-shaped torus ($\gamma > 1)$, (d) horn torus without a hole ($\gamma=1$), and (e) spindle torus ($0 < \gamma < 1$ (the shape of half a lemon is visible inside). Picture taken from \cite{Gabbrielli_2014}, for illustrative purposes.
}
\end{figure}

In the current work we shall not delve into a detailed description of the compactification scheme, as this will not be necessary for our purposes. Instead we shall make use of a basic feature of toroidal compactification in a simple scheme in which the six-dimensional manifold is represented as a product of three tori 
\begin{align}\label{sixspace}
{\rm 6-dimensional~compact~space} = T^2_{1} \otimes T^{2}_2 \otimes T^2_{3}\,.
\end{align}
Each torus is a two-dimensional manifold, topologically equivalent to a product of circles  $T^2 \sim S^1 \otimes S^1$, which can be visualized as a compactified cylindrical tube, \emph{i.e.} a cylinder with its ends identified. This leads to the existence of minor (transverse cross section of the tube) and major (along the compactified tube surface) circles
with radii, say, $R_1 \lesssim R_2$, 
respectively. In string theory there is no special  preference for the ratio
\begin{align}\label{ratioR}
\gamma  \equiv \frac{R_1}{R_2}\,, 
\end{align}
which can take any real value. 
This value determines the shape of the torus~\cite{Gabbrielli_2014} (\emph{cf.} Fig.~\ref{fig:tori}). For instance, the case $\gamma > 1 $ corresponds to the standard (ring) torus, a traditional "doughnut"(or ``donut'')-shaped manifold with a hole in the middle. The degenerate case 
\begin{align}\label{horn}
\gamma_{\,{\rm horn-torus}} = 1\,, \qquad (R_1 = R_2) \,,
\end{align}
corresponds to 
the so called horn-torus, according to the classification and nomenclature of \cite{Gabbrielli_2014}, which is a two-dimensional toroidal manifold without a hole.

A ``flat'' torus is constructed from a rectangle with its opposite sides identified (``glued''). It is customary to express the ratio of the side lengths as a complex ``moduli'' parameter $\tau$. Although a torus has a specific topology, it can be characterized, as we have seen above, by different shapes, associated, \emph{e.g,} with the sizes of the radii $R_i, i=1,2$. These shapes define the ``moduli'', classified by the parameter $\tau$. 
For instance, in the case the rectangle is a square, we have 
\begin{align}\label{square}
\tau=i\,. 
\end{align}
The value 
\begin{align}\label{hex}
\tau=\exp (i\pi/3) = \frac{1}{2} + i \frac{\sqrt{3}}{2}\
\end{align}
corresponds to a ``hexagonal'' torus. The latter is constructed out of a hexagonal tilling of the infinite plane by identifying in a given finite patch opposite edges in two different directions: one pair of parallel directions, and another pair rotated by $60^o$ with respect to the first. In contrast to the square torus, in which each point has four nearest neighbours, in the hexagonal torus, each point has six nearest neighbours, due to the underlying honeycomb-lattice structure of space. 

The hexagonal torus is built from equal-length cycles meeting at $60^o$, with 
the respective trigonometry of that angle forcing the ratio of the effective major to minor radii to be 
\begin{align}\label{hexgamma}
\gamma_{\,\rm hex-torus}=\sqrt{3}\,,
\end{align}
if the torus is embedded in $\mathbb R^3$ as a doughnut.
The hexagonal torus has the maximal symmetry among flast tori, a 6-fold rotational symmetry by $60^o$ and the largest automorphism group of any flat torus.
In order to visualize the corresponding geometry, we mention that such a geometry underlines Graphene. The moduli parameter \eqref{hex} constitutes a fixed point of the modular group of transformations 
$ \tau \rightarrow \frac{a \tau + b}{c \tau + d}\,,$ with $ a\, d - b\, c = 1\,$, specifically of $\tau \rightarrow - \frac{1}{\tau + 1}$, using modular equivalence $\tau \sim \tau -1$.

It is important to remark that the hexagonal torus \eqref{hex}, characterised by the radii-ratio \eqref{hexgamma} plays an important r\^ole in string compactification, in particular orbifold compactification (see, \emph{e.g.} \cite{Forger:1997tu,Becker:2006dvp,Fischer:2012qj,Chattopadhyaya:2016xpa,Larotonda:2026hxy}).
The radii in the above discussion play the r\^ole of additional infrared cutoffs in the approach of \cite{Sanchis-Lozano:2022nzp,sanchis-sanz:2024}, on top of the one associated with spectrum of the inflaton field. 
In what follows we shall consider the cases (\emph{cf.} Fig.~\ref{fig:tori}) of hexagonal \eqref{hexgamma} and horn tori \eqref{horn}, 
in an attempt to provide the best fit to the CM data temperature and polarization anisotropy data at small-multipoles. 

Our interest in this work will concern orbifold-type toroidal compactification, $T^6/(Z_2 \otimes Z_2)$, which can be thought of as implying that each of the extra torus dimensions will be represented as a line segment, at each end of which appropriate boundary conditions of even and odd parity for a scalar field assumed to be present in the early Universe (\emph{e.g.} the inflaton of standard cosmologies) will be imposed. In other words we shall replace each torus $T^2$ in $T^6$ by $T^2/Z^2$. 
Let us first consider the cases of a torus $T^{(2)}/Z_2 \sim (S_1 \otimes S_1)/Z_2$ for which the radii of the two circles $S_1$ are the \emph{same}. We shall consider the general case in which the circles have different radii later on. 

\subsection{Isotropic (Horn) Tori compactification}\label{sec:horn}

Let us assume that each torus in $T^6$ is characterized by a radius scale $R_i, \, i=1,2,3$. 
We start by discussing the most general case where these radii are in general different. 
If we denote $(x_\mu, z^i)$, $\mu=0, \dots 3$, $i=1,2,3$, 
the ten dimensional coordinates of the target spacetime of the string (assuming superstrings living in their critical dimension)~\cite{Green:2012oqa}), where $x^\mu \in \mathbb R$ are the uncompactified coordinates, while $z^i \in \mathbb C$ denote the complex toroidal coordinates in each of the three tori $T^2$ comprising \eqref{sixspace}, then, for any field, and hence for our inflaton $\phi (x^\mu, z^i) \equiv \phi (x, z^i)$, propagating in the extra dimensions as well, one would have, under the action of the orbifold parity, that is the $Z_2$ action action 
\begin{align}\label{z2}
z^i \to - z^i\,,
\end{align}
the following behaviour of the field:
\begin{align}\label{oriparfield}
\phi (x, -z^i) = {\mathcal P}_i \, \phi(x, z^i)\
\end{align}
where 
\begin{align}
P_i = +1 \rightarrow {\rm Parity~even}\ \ \ ;\ \ 
P_i = -1 \rightarrow {\rm Parity~odd}\ \ \ \label{odd}\,.
\end{align}

If the scalar field is subject to a $Z_2$ symmetry,
it can be decomposed into even and odd components,
$\phi_{+}$ and $\phi_{-}$, respectively. The Kaluza--Klein
expansions along the compact directions take the form

\begin{equation}
\phi_{+}(x,y)
=
\sum_{\{n_A\}}
\phi^{(+)}_{\{n_A\}}(x)
\prod_{A=1}^{6}
\cos\!\left(\frac{n_A}{R_A}y^A\right), \qquad ({\rm even~parity})\,,
\end{equation}

\begin{equation}
\phi_{-}(x,y)
=
\sum_{\{n_A\}}
\phi^{(-)}_{\{n_A\}}(x)
\prod_{A=1}^{6}
\sin\!\left(\frac{n_A}{R_A}y^A\right) \,, \qquad ({\rm odd~parity})\,.
\end{equation}

Here $n_A$ denote the Kaluza--Klein mode numbers associated
with the six compact circles $S^1_{R_A}$ and $R_A$ are the
corresponding radii. The cosine modes correspond to fields that are even under
the $Z_2$ symmetry and therefore admit zero modes, while
the sine modes correspond to odd fields whose zero modes
are projected out.

The KK modes  mass spectra 
are given in terms of the (discrete) momentum numbers $m,n$
in each $T^{(2)}_i$, $i=1,2,3$, as
\begin{equation}\label{KKmassspectrum}
M^2 = \sum_{i=1}^{3} \frac{n_i^2 + m_i^2}{R_i^2}\,.
\end{equation}

\subsubsection{The case of one extra dimension and IR cutoffs, as a warm up}

Before proceeding with the KK spectrum, and the relevance t our IR cutoff discussion here, let us first review the instructive case of one extra dimensional case of \cite{sanchis-sanz:2024,Sanchis-Lozano:2025csn} where the extra dimension is a circle of radius $R$, but the segment is orbifoldised, that is, it is 
viewed as a \emph{fundamental} region of a compact coordinate $\mathcal Y$ with boundaries at the fixed points $0$ and $\pi R$, then the even and odd parity boundary conditions in each $T^2_i$, $i=1,2,3$, can be replaced by Neumann or Dirichlet boundary conditions for the scalar field.  This actually is what happening in each of the factors of the $T^6/(Z_2\otimes Z_2)$ compactification, if we 
parametrise each $T^2$ as a product of two circles, $S_1$, $T^2 = S_1 \otimes S_1$, and orbifoldise each $S_1$, by viewing it as a segment of a single compact dimension stretching between the fixed points $\mathcal Y \in [0, \pi R_A]$, $A=1, \dots 6$, where the index $A$ runs over the $S_1$'s. 

Specifically, odd-parity boundary conditions 
for the field \eqref{oriparfield} and \eqref{odd}, 
correspond to Dirichlet conditions at the fixed points:
\begin{align}\label{dirichlet}
    \phi(x,\mathcal Y) = -\phi(x,-\mathcal Y)  \quad  \Rightarrow \quad  \phi(x,0)=\phi(x, \pi R) =0\,,
\end{align}
while even-parity boundary conditions correspond to Neumann boundary conditions at the fixed points:
\begin{align}\label{neumann}
    \phi(x,\mathcal Y) = \phi(x,-\mathcal Y)  \quad  \Rightarrow \quad  \phi(x,0)=\phi(x, \pi R) \quad \Rightarrow \quad \partial_{\mathcal Y} \phi \Big|_{\mathcal Y = 0, \, \pi R} =0 \,.
\end{align}
In such a case, of either Dirichlet or Neumann boundary conditions at the two ends of the compact dimension, one obtains a closely related KK mass spectrum as in the odd, even parity cases, with $M_n^2 = \frac{n^2}{R^2}$, with $n \ge 0$ for Neumann (allowing for massless scalar fields), 
and $n \ge 1$ for Dirichlet boundary conditions (no massless states allowed). 

An interesting case, which as we shall see, provides the best fit to the cosmological data, specifically temperature and polarization anisotropies of the CMB, is the one in which one imposes \emph{mixed} boundary conditions at the fixed points of the fundamental domain of the compact variable $\mathcal Y$: 
\begin{align}\label{mixed}
\phi (x, 0) = 0, \qquad \partial_{\mathcal Y} \phi\Big |_{\mathcal Y = \pi R}= 0 \qquad ({\rm and~vice~versa})\,.
\end{align}
The corresponding KK mass spectrum in this case is half-shifted, as it is called: 
$M_n^2 = \frac{(n + \frac{1}{2})^2}{R^2}$.

For $T^6/(Z_2 \otimes Z_2)$, summarizing the results for all three of the above cases: 
\begin{table}[ht]
\centering
\begin{tabular}{cccc}
\toprule
Boundary condition & Boundary constraints & KK wavefunctions & Mass spectrum \\ 
\midrule

Neumann (N,N) &
$\partial_{\mathcal Y} \phi\Big |_{\mathcal Y = 0,\,\pi R}=0$ &
$\cos\!\left(\dfrac{n \mathcal Y}{R}\right)$ &
$M_n^2=\dfrac{n^2}{R^2},\quad n=0,1,2,\dots$ \\

Dirichlet (D,D) &
$\phi(0)=\phi(\pi R)=0$ &
$\sin\!\left(\dfrac{n \mathcal Y}{R}\right)$ &
$M_n^2=\dfrac{n^2}{R^2},\quad n=1,2,3,\dots$ \\

Mixed (D,N) &
$\phi(0)=0,\quad \partial_{\mathcal Y}\phi|_{\pi R}=0$ &
$\sin\!\left(\dfrac{(n+\tfrac12)\mathcal{Y}}{R}\right)$ &
$M_n^2=\dfrac{(n+\tfrac12)^2}{R^2},\quad n=0,1,2,\dots$ \\

Mixed (N,D) &
$\partial_{\mathcal Y}\phi|_{0}=0,\quad \phi(\pi R)=0$ &
$\cos\!\left(\dfrac{(n+\tfrac12)\mathcal Y}{R}\right)$ &
$M_n^2=\dfrac{(n+\tfrac12)^2}{R^2},\quad n=0,1,2,\dots$ \\

\bottomrule
\end{tabular}
\caption{Kaluza--Klein wavefunctions and mass spectra for a scalar field
on the fundamental  interval of a compact dimension  $0\le \mathcal Y\le \pi R$ with Neumann (N), Dirichlet (D), and mixed
boundary conditions.}
\label{BC_DNM_KK_spectrum}
\end{table}

Thus, for the case of a single extra compact dimension, studied in 
\cite{sanchis-sanz:2024,Sanchis-Lozano:2025csn}, we have for the ratio of masses 
\begin{align}\label{massratio}
\frac{M_n^{(D,D)}}{M_n^{(N,D)}} = \frac{n + 1}{n + \frac{1}{2}} \stackrel{n=0}{=} 2 \,, \quad n = 0, 1, \dots \,,
\end{align}
implying a ratio 2 when $n=0$, which is the case providing the IR cutoff for the scalars in (3+1) uncompactified spacetime dimensions. So the corresponding IR cutoffs $k^{(D,D)}$ and $k^{(N,D)}$ are related by a factor of 2. This is essential for providing the best fit to temperature and polarization CMB data as we shall see.

\subsubsection{KK mass spectrum in the $T^6/(Z_2\otimes Z_2)$ case}

Extension to the $T^6/(Z_2 \otimes Z_2)$ case is straightforward, yielding the KK mass spectrum:
\begin{equation}
M^2 = \sum_{i=1}^{3} \frac{\kappa_i^2}{R_i^2}, \qquad
\kappa_i =
\begin{cases}
n_i, & \text{Neumann} \\[2mm]
n_i, & \text{Dirichlet } (n_i \ge 1) \quad .\\[1mm]
n_i + \frac{1}{2}, & \text{Mixed}
\end{cases}
\end{equation}
Hence, the Neumann boundary condition allows for a zero mode, whilst the Dirichlet one removes it. The Mixed boundary conditions, on the other hand, produce half-integer KK modes
$n+ \frac{1}{2}$, $n =0, 1, 2 \dots $, called to play a crucial role in our develepment.

As already mentioned, we consider specific toroidal $T^6$ compactification, characterised by a common ratio $\gamma$  \eqref{ratioR} of the radii in all three tori $T^2_i$, $i=1,2,3$, comprising $T^6$, see  Eq.~\eqref{sixspace}. We also assume that the three tori in $T^6$ are identical copies of each other, as far as imposition of the boundary conditions is concerned. This specifies the compactification scheme. 

Following the analysis leading to \eqref{massratio}, then, we also conclude that in this more general case, one can obtain a ratio 2 between the pertinent IR cutoffs corresponding to (even,odd) parity (or (N,D) boundary conditions) modes of the inflaton field.

\subsection{Anisotropic Six--Torus Compactification}
\label{sec:T6_anisotropic}

Finally, we proceed to discuss the anistrotopic $T^6$-orbifold case, in which each tori is represented as a product of orbifoldised circles, $T^2 \sim S^1 \otimes S^1$, with different radii $R_{1,2}$, characterised by a ratio \eqref{ratioR}. In the beginning we shall be fairly general, but for our purposes here we shall eventually restrict ourselves to the identical-copies scheme for each of the three tori of $T^6$, for concreteness and convenience.

In the anisotropic case the six circles are allowed to have
different radii. We therefore introduce six compact coordinates
\( y^A \) with
\( A=1,\dots,6\).

\subsubsection{The Geometry}

The compact coordinates satisfy the periodic identifications

\begin{equation}
y^A \sim y^A + 2\pi R_A ,
\qquad A=1,\dots,6 .
\end{equation}

The metric of the compact space can be written as

\begin{equation}
ds^2 = \sum_{A=1}^{6} (dy^A)^2 .
\end{equation}

The total volume of the six--torus is therefore

\begin{equation}
V_6 = \prod_{A=1}^{6} (2\pi R_A)
    = (2\pi)^6 \prod_{A=1}^{6} R_A .
\end{equation}

We next proceed to discuss the KK spectrum. 
\subsubsection{Kaluza--Klein Expansion ands Mass spectrum}

Consider a scalar field $\phi(x,y)$ propagating on
$M^4 \otimes T^6$, where $M^4$ is the (3+1)-dimensional spacetime, which in the case of the early Universe describes the appropriate cosmological time. For our purposes here we ignore details of the cosmological expansion, and assume the space $M^4$ flat, for all practical purposes. This approximation suffices, as the details of the Universe (during its inflationary phase) will not play an important r\^ole for our purposes. 

The field can be expanded in Fourier modes
along the compact directions,

\begin{equation}
\phi(x,y)
=
\sum_{\{n_A\}}
\phi_{\{n_A\}}(x)
\exp\left(
i\sum_{A=1}^{6}
\frac{n_A}{R_A} y^A
\right),
\end{equation}

where $n_A$ denote the Kaluza--Klein excitation numbers
associated with the six compact circles.

The four--dimensional mass of a Kaluza--Klein mode is determined by
the quantized momentum in the compact directions,

\begin{equation}
M^2 =
\sum_{A=1}^{6}
\frac{\kappa_A^2}{R_A^2}.
\label{KKmass}
\end{equation}

The allowed values of $\kappa_A$ depend on the boundary conditions
imposed on the scalar field along the corresponding circle,

\begin{equation}\label{kabc}
\kappa_A =
\begin{cases}
n_A, & \text{Neumann boundary conditions}, \\[6pt]
n_A, & \text{Dirichlet boundary conditions} \quad (n_A \ge 1), \\[6pt]
n_A + \dfrac{1}{2}, & \text{mixed boundary conditions}.
\end{cases}
\end{equation}

The anisotropy of the six--torus leads to several important
features:

\textbf{(i)} The Kaluza--Klein spectrum is generally non--degenerate since
the radii $R_A$ may differ along the six compact directions. Degeneracies will occur when we consider the special case in which the $T^6$ is the product of three identical copies $T^2$.

\textbf{(ii)} The characteristic mass scale of Kaluza--Klein excitations
associated with the circle $S^1_{R_A}$ is

\begin{equation}
M_{\mathrm{KK}}^{(A)} \sim \frac{1}{R_A}.
\end{equation}

\textbf{(iii)} Zero modes exist only when Neumann boundary conditions are
imposed along the corresponding compact directions.

\textbf{(v)} For the zero mode $n=0$ we can obtain  again, as in the single extra compact dimensional case of \cite{sanchis-sanz:2024,Sanchis-Lozano:2025csn} (\emph{cf.} \eqref{massratio}), 
the ratio 2 between the pertinent  the pertinent IR cutoffs in the CMB temperature and polarization anisotropies.

For a scalar field compactified on an anisotropic six--torus,
the four--dimensional Kaluza--Klein spectrum is determined by

\begin{equation}\label{M2s6}
M^2 =
\sum_{A=1}^{6}
\frac{\kappa_A^2}{R_A^2},
\end{equation}
where $R_A$ denote the radii of the six compact circles and
the allowed mode numbers $\kappa_A$ depend on the boundary
conditions imposed along each circle.

It is convenient for our purposes to rewrite \eqref{M2s6} as a double sum, as follows:
\begin{equation}\label{M2kij}
M^2 = \sum_{A=1}^{6}
\frac{\kappa_A^2}{R_A^2} = 
\sum_{i=1}^{3}\sum_{I=1}^{2}
\frac{\kappa_{iI}^2}{R_{iI}^2}\,, \end{equation}
where $i=1, 2,3$ labels the tori $T^2$ in the product corresponding to $T^6$, and $I=1,2$ labels the radii inside each torus $T^2_i$, $i=1,2,3$, with $I=1$ corresponding to the ``exterior'' radius, as appearing in the ratio $\gamma$ (\emph{cf.} \eqref{ratioR}). The quantities $\kappa_{iI}$ are zero, positive integers or positive half integers, as follows from the boundary conditions \eqref{kabc}.

In our case of interest, we shall consider the two cases, horn and hexagonal torus, with ratios of radii in each $T^2$, equal to $\gamma =1, \sqrt{3}$, respectively (\emph{cf.} \eqref{horn}, \eqref{hexgamma}).
We shall see that the data show a slight preference to the horn torus case, as compared to the hexagonal torus, but of course, due to the current large errors in the data, the situation is inconclusive. For simplicity and concreteness, we assume compactifications in which $\gamma$ is common among the three tori $T^{(2)}$ in $T^6$, and also the radii 
among ``successive'' tori (assuming a given ordering in the product $T^6 = T^{2}_1 \otimes T^{2}_2 \otimes T^{2}_3$) have a ratio 2, that is we have the following ratios among $R_{i1}$, $i=1,2,3$:
\begin{align}\label{radiitori}
\frac{R_{i1}}{R_{i2}} &= \gamma = 1, \, \sqrt{3}\, \qquad i=1,2,3\,, \nonumber \\
\frac{R_{11}}{R_{21}} &=\frac{R_{21}}{R_{31}} = 2\,.
\end{align}
In this notation, \eqref{M2kij} is written as:
\begin{align}\label{m2Rij}
M^2 = \frac{1}{R_{11}^2} \Big(\kappa_{11}^2 + 4\,\kappa_{21}^2 + 16\, \kappa_{31}^2 + \gamma^2 \Big[\kappa_{12}^2 + 4\,\kappa_{22}^2 + 16\, \kappa_{32}^2 \Big] \Big)\,.
\end{align}
Different combinations of the lowest states shown in the above expression determine the IR cutoff assignments required to fit the angular correlation data at large scales in our later analysis. On the other hand, the absolute scale of the $T^2$ radii, set by $R_{11}$, will be determined in accordance with the physics underlying the CMB anisotropies. 

Before starting our phenomenological analysis of the CMB anisotropies, let us introduce an important caveat. Incorporating extra spatial dimensions into a realistic four-dimensional cosmology is highly non-trivial, especially once inflation is taken into account. Building a consistent framework that simultaneously accommodates stable compactification, viable inflation, and standard post-inflationary cosmology remains one of the main challenges in theories with extra dimensions \cite{Lidsey:1999mc}. In this work, however, we do not attempt to address these fundamental issues; instead focus on their potential observable consequences,  interpreted within this still open theoretical framework.

Even if we are currently far away from observational sensitivities that could realistically constrain the topology of the string extra-dimensional spacetimes in the CMB spectrum, nonetheless we hope that our current contribution will prove useful to guiding realistic compactifications in string theory in the future, when (and if) the observational technologies become sufficiently advanced to allow for accurate measurements (of angular correlations in the sky) in the low-multipole part of the temperature angular power spectrum of the CMB.

\section{CMB angular correlations revisited}\label{sec:angspec}

Having outlined the essential elements of the compactification framework adopted in this work, we first revisit earlier studies that highlight its basic features~\cite{sanchis-sanz:2024}, together with the key assumptions and procedures employed in the prior analysis of Planck data~\cite{Planck:2015gmu,Planck:2019evm}, before proceeding to the new study of CMB angular correlations.

One of the most remarkable results from measurements of the CMB temperature is its extraordinary homogeneity across the full sky, implying a high degree of isotropy in all directions. Nevertheless, small but significant deviations from perfect uniformity encode valuable information about the early Universe. In this context, correlations between temperature anisotropies measured at different points on the sky depend only on their angular separation. Consequently, the two-point angular correlation function provides a natural and powerful tool to characterize these fluctuations.

It is defined as the ensemble average of the product of the CMB temperature fluctuations, $\delta T(\hat{n})$, relative to the mean temperature, evaluated over all pairs of directions on the sky specified by the unit vectors $\hat{n}_1$ and $\hat{n}_2$:
\begin{equation}\label{eq:CTT}
C^{\rm TT}(\theta)=\langle \delta T(\hat{n}_1)\delta T(\hat{n}_2) \rangle \;,
\end{equation}
where $\theta \in [0,\pi]$ is the angle between the two directions, defined through $\hat{n}_1 \cdot \hat{n}_2 = \cos\theta$.

Expanding the temperature fluctuations in spherical harmonics, $Y_{\ell m}(\hat{n})$, statistical isotropy implies that, after averaging over the azimuthal angle, the angular power spectrum coefficients $C_\ell$ depend only on $\ell$. Consequently, the auto-correlation function $C^{\rm TT}(\theta)$ depends only on the angular separation $\theta$ and admits an expansion in Legendre polynomials:
\begin{equation}\label{eq:C2}
C^{\rm TT}(\theta) = \sum_{\ell \ge 2} \frac{2\ell + 1}{4\pi} C_{\ell}^{\rm TT} P_{\ell}(\cos{\theta})\;,
\end{equation}
where $P_{\ell}(\cos{\theta})$ denotes the Legendre polynomial of order $\ell$. The sum formally runs from $\ell = 2$, since the monopole and dipole contributions are removed, to infinity. In practice, truncating the series at $\ell \sim 200$ is sufficient.

 As alredy mentioned, the expansion \eqref{eq:C2} relies on the assumption of statistical isotropy, which ensures that the covariance of the spherical harmonic coefficients is diagonal and independent of the azimuthal index. This, in turn, implies that the correlation function depends only on the angular separation $\theta$. However, this assumption does not enforce parity symmetry in the sense used in this work. In particular, an asymmetry between even- and odd-$\ell$ multipoles, as we shall see, corresponds to a violation of parity invariance but actually does not imply any preferred direction on the sky. Therefore, the observed odd-parity preference can be consistently accommodated within a statistically isotropic framework.

On the other hand, the $\ell$-mode can be roughly related to the angle through the simple expression $\theta \sim \pi/\ell$, so that small $\ell$ values have a greater impact at large angles, say above $90^{\circ}$. Nonetheless, despite this association, the influence of the Legendre polynomial $P_{\ell}(\cos{\theta})$ extends over the entire angular range under study. Therefore, the two-point angular correlation function and the primordial power spectrum play complementary roles in the analysis carried out in this paper.

In this work, we restrict our study to (self) two-point correlation functions. Higher-order correlators have proven instrumental in testing inflationary non-Gaussianities~\cite{Chen:2006nt,Seery:2005gb,Vernizzi:2006ve,Ellis:2014rja}, and could provide important additional constraints in our framework. Moreover, cross-correlations, despite their relevance, are not considered here; in particular, TB and EB correlations arise only in the presence of parity-violating physics~\cite{Gluscevic:2010vv}. These aspects will be addressed in future work.

\subsection{Scalar and tensor primordial power spectra}\label{sec:scaltens}

In cosmological perturbation theory, scalar, tensor, and vector modes contribute differently to CMB temperature anisotropies. Scalar modes, arising from primordial density fluctuations, dominate across all scales, producing the Sachs–Wolfe plateau at low $\ell$ and the acoustic peaks at intermediate $\ell$. Tensor modes, corresponding to inflationary gravitational waves, primarily affect large angular scales ($\ell \lesssim 100$) and are subdominant in the temperature spectrum. Vector modes, which describe vortical perturbations, are typically negligible unless sourced by non-standard mechanisms such as topological defects or anisotropic stresses. 

In this work, we neglect vector modes and focus on scalar and tensor contributions, which will be addressed below by recalling some fundamental aspects of their primordial power spectra, with a particular emphasis on the emergence of infrared cutoffs from EDs.

\subsubsection{Scalar power spectrum}

In studies of CMB temperature anisotropies, the primordial power spectrum characterizes the statistical distribution of the initial cosmological perturbations generated in the early Universe, typically during inflation. It is defined in Fourier space as the variance of the primordial curvature (or density) perturbation amplitudes per logarithmic interval in comoving wavenumber $k$. In the simplest models, where the spectrum is nearly scale invariant, it can be parametrized as 
\begin{equation*}\label{eq:spower}
    {\cal P}^S(k) = A_s \biggl( \frac{k}{k_{\ast}} \biggr)^{n_s-1}
\end{equation*}
where $A_s$ and $n_s$ denote the amplitude and the spectral index, respectively, and $k_{\ast}$ the pivot scale which in the case of {\it Planck} is taken as $k_{\ast} = 0.05~ {\rm Mpc}^{-1}$. 

The temperature contrast is expanded in spherical harmonics, and the angular power spectrum is defined as $D_{\ell}=\ell(\ell+1)/2\pi\ C_{\ell}$ where denotes the  coefficients of the multipole expansion function. Under the assumptions of statistical isotropy and Gaussianity, the power spectrum fully encodes the two-point correlation properties of the anisotropies. It therefore constitutes a primary observable for confronting theoretical models with observations, constraining cosmological parameters.

\subsubsection{Tensor power spectrum}\label{sectensorp}

For tensor modes, an expression analogous to Eq.~\eqref{eq:spower} applies:
\begin{equation}\label{eq:tpower}
{\cal P}^T(k) = A_t \left( \frac{k}{k_{\ast}} \right)^{n_t}
\end{equation}
where, for historical reasons, the spectral index is defined without the $-1$ offset.

The relative weight of tensor to scalar modes is conveniently parametrized by the ratio $r(k_{\ast}) = P^T/P^S$. This quantity plays a fundamental role in CMB polarization; in our case, it is also of particular importance for the analysis of temperature anisotropies at large angular scales.

In scenarios where the fundamental gravitational theory is formulated in higher dimensions, the relation between the primordial tensor spectrum and the observable CMB tensor contribution may differ from the conventional four-dimensional case. To explore this possibility, we assume that the tensor power spectrum factorizes into multiple contributions (corresponding to three toroidal sectors), each characterized by its own infrared cutoff doublet, following Eq.~\eqref{radiitori}. 

These contributions are expected to affect the CMB temperature and polarization spectra, as discussed in the following sections. 


Assuming a highly symmetric compactification, as discussed in section \ref{sec:torcomp}, in which the internal sectors contribute with comparable normalization, the tensor spectrum can be approximated as
\begin{equation}\label{eq:sumP}
{\cal P}^{\rm T}(k) \to\ \alpha(k)\ {\cal P}^{\rm T}(k)\;,
\end{equation}
where $\alpha(k)$ mainly enhances low $k$ contributions, and therefore low multipoles:
\begin{eqnarray}\label{eq:alpha}
    \alpha(k) & \simeq & {\cal O}(1-10),\ k\ \lesssim k_{\rm min}\,, \qquad {\rm for~tori~compactification}\,, \nonumber \\ \alpha(k) & \simeq & 1\ ,\ k > k_{\rm min}\,.
\end{eqnarray}

The enhancement factor $\alpha$ can be interpreted as arising from the cumulative contribution of Kaluza–Klein tensor modes associated with the compact extra dimensions. {\color{black} In simple toroidal compactifications, the number of relevant internal sectors is expected to be of the order of the number of compact dimensions, while the overall normalization is left as a phenomenological parameter. So we tentatively write}
\begin{equation}\label{eq:alpha2}
\alpha \sim {\cal O}(1-10)\; ,
\end{equation}
{\color{black} to be numerically determined from the observed power spectrum combining both scalar and tensor modes. Therefore $\alpha \sim 10$ is not derived from a complete higher-dimensional calculation; rather it can be seen as a phenomenological effective parameter accounting for the bulk tensor contributions relative to the effective four-dimensional power spectrum. 

The resulting enhancement is expected to affect predominantly the largest angular scales, while leaving the spectrum at smaller scales essentially unchanged. This is also consistent with the standard evolution of inflationary tensor perturbations. At recombination, large-scale tensor modes (small $k$) enter the horizon later and can therefore contribute more significantly to the lowest multipoles. By contrast, smaller-scale modes (larger $k$) enter the horizon earlier and undergo a stronger suppression, making their contribution to the CMB anisotropies increasingly negligible at high multipoles, in particular for $\ell\gtrsim100$ \cite{Annis:2022xgg}.}

Actually, the decrease of $\alpha(k)$ should not be abrupt, but smoothed according to 
\begin{equation*}
    \alpha(k) = 1 + (\alpha_0-1)\ \exp(-k/k_{\rm min})^p
\end{equation*}
where $\alpha_0 \simeq$ ${\cal O}(1-10)$; $p \geq 1$ is a shape (steepness) parameter controlling the smoothness of the transition. For definiteness, we adopt $p=1$ in later calculations. 

Assuming the approximate relation $k \simeq \ell / r_L$, where $r_L$ denotes the comoving distance to the last-scattering surface (LSS), and taking $k_{\rm min} \simeq {\rm few} \times 10^{-4}\ \mathrm{Mpc}^{-1}$ \cite{Melia:2018,Sanchis-Lozano:2022zrp}, the affected multipoles satisfy $\ell \lesssim 30$. This corresponds to the large-scale regime on which we focus. 

The ratio of tensor to scalar multipole coefficients becomes
\begin{equation}
\frac{C_\ell^{(T)}}{C_\ell^{(S)}}\ \simeq\ \alpha(\ell)\ r(k_{\ast})\ ;\ \ \alpha(\ell) = 1 + (\alpha_0-1)\ \exp(-\ell/\ell_{\rm min})
\end{equation}
with $ \ell_{\rm min} = k_{\rm min} r_L$ ,
representing a modest but significant enhancement of the low multipole coefficients for $\ell \lesssim 30$, while remaining consistent with the present bound $r(k_{\ast}) < 0.034$~\cite{Planck:2018jri,Planck:2018vyg}. Note that this enhancement mechanism applies to all IR cutoffs arising from compactification, as will be discussed below.


\subsection{Infrared cutoff in the scalar power spectrum}

Under some simplifying assumptions and for angles larger than a few degrees \cite{Mukhanov:2005}, the coefficients $C_{\ell}$ in Eq~\eqref{eq:C2} can be evaluated through the Sachs-Wolfe (SW) effect according to
the following expression 
\begin{equation}\label{eq:Cell}
C_{\ell}^{\rm TT}\ =\ N\ \int_{0}^{\infty}dk\ k^{n_s-2}\ j_{\ell}^2(k\ r_L)\ \to C_{\ell}^{TT}\ =\ N_S\ \int_{0}^{\infty}du\ \frac{j_{\ell}^2(u)}{u}\;,
\end{equation}
where $N_S$ is a normalization constant, $n_s \approx 1$ is the spectral index and $j_{\ell}(k\ r_{d})$ denotes the spherical Bessel function
of order $\ell$, whose argument involves the comoving distance $r_L$ to the LSS (typically about 14,000 Mpc), and the wavenumber $k$ for each mode in the CMB power spectrum 
which varies, in principle, from zero to infinity.

However, focusing on the two-particle angular  correlation function, a very poor agreement between theory and observations emerges \cite{Melia:2018,Melia:2021tvx}. To address this unsatisfactory situation, the authors introduced an IR cutoff $k_{\min}$ into the scalar power spectrum, which in turn implied a lower limit $u_{\min}$ in the integral defining the $C_{\ell}$ coefficients, as shown below:
\begin{equation}\label{eq:Cellcutoff}
C_{\ell}^{\rm TT}{\rm (scalar)}\ =\ N_S\ \int_{u_{\rm min}}^\infty\ du\ \frac{j_{\ell}^2(u)}{u}\;,
\end{equation}
where $u_{\rm min}$ is related to the IR cutoff as
\begin{equation}\label{eq:u1}
k_{\rm min}= \frac{u_{\rm min}}{r_L}\;,
\end{equation}
 and the reader should bear in mind that $r_L$, which  denotes the co-moving distance to the LSS, sets a causality scale limit to angular correlations across the sky \cite{Hogan:2021pap}.  

  Notice that $k_{\rm min}$ (and therefore the corresponding lower dimensionless limit $u_{\rm min}$) denotes a sharp cutoff in the scalar power spectrum (later extended to the tensor power spectrum), which removes all Fourier modes below this scale. The idea of truncating the primordial power spectrum was originally proposed in \cite{Contaldi:2003zv}, and later extensively applied to CMB angular correlations in a long series of works \cite{Melia:2018,Melia:2021tvx,Liu:2025yvp}.

The main effect of such a cutoff is to significantly suppress the contribution of the lowest multipole modes to $C^{\rm TT}(\theta)$, particularly the quadrupole term, thereby inducing a certain degree of parity breaking as discussed in \cite{Sanchis-Lozano:2022nzp}.  However, the impact of only one or two multipoles is not sufficient to account for the long-standing odd-parity preference observed in CMB analyses. {\color{black} A similar conclusion can be drawn from the so-called Ellipsoidal Universe \cite{Cea:2022zep}, in which the flat Friedmann--Lemaître--Robertson--Walker metric is replaced by an anisotropic metric that significantly affects only the quadrupole term.}

Consequently, the overall agreement between theoretical predictions and observations remains unsatisfactory, as illustrated in Fig.~\ref{fig:3}, indicating that additional mechanisms are required.
 
\subsection{Double infrared cutoff in the scalar power spectrum}

The original idea of introducing a double-cutoff structure in the scalar power spectrum was proposed in \cite{Sanchis-Lozano:2022nzp} to improve the fit of the two-point correlation function $C^{\rm TT}(\theta)$ to {\em Planck} data~\cite{Planck:2018vyg} over the angular range $[4^{\circ},180^{\circ}]$, with particular attention to the large angular region mainly governed by low multipoles in the Legendre expansion \eqref{eq:C2}. Indeed, the combined effect of two different lower cutoffs for odd and even multipoles enhances the parity imbalance indicated by observational data.

In order to facilitate the discussion, we split $C^{\rm TT}(\theta)$ into two contributions corresponding to even and odd multipoles,
\begin{equation}\label{eq:Csplit}
C(\theta)^{\rm TT}= C_{\rm even}^{\rm TT}(\theta)+C_{\rm odd}^{\rm TT}(\theta)=\frac{1}{4\pi}\sum_{\ell_{\rm even}} (2\ell+1)\ C_{\ell}^{\rm TT}\ P_{\ell}(\cos\theta)
+\frac{1}{4\pi}\sum_{\ell_{\rm odd}}(2\ell+1)\ C_{\ell}^{\rm TT}\ P_{\ell}(\cos{\theta}).
\end{equation}

\begin{figure}

\centering
\includegraphics[width=7.0cm]{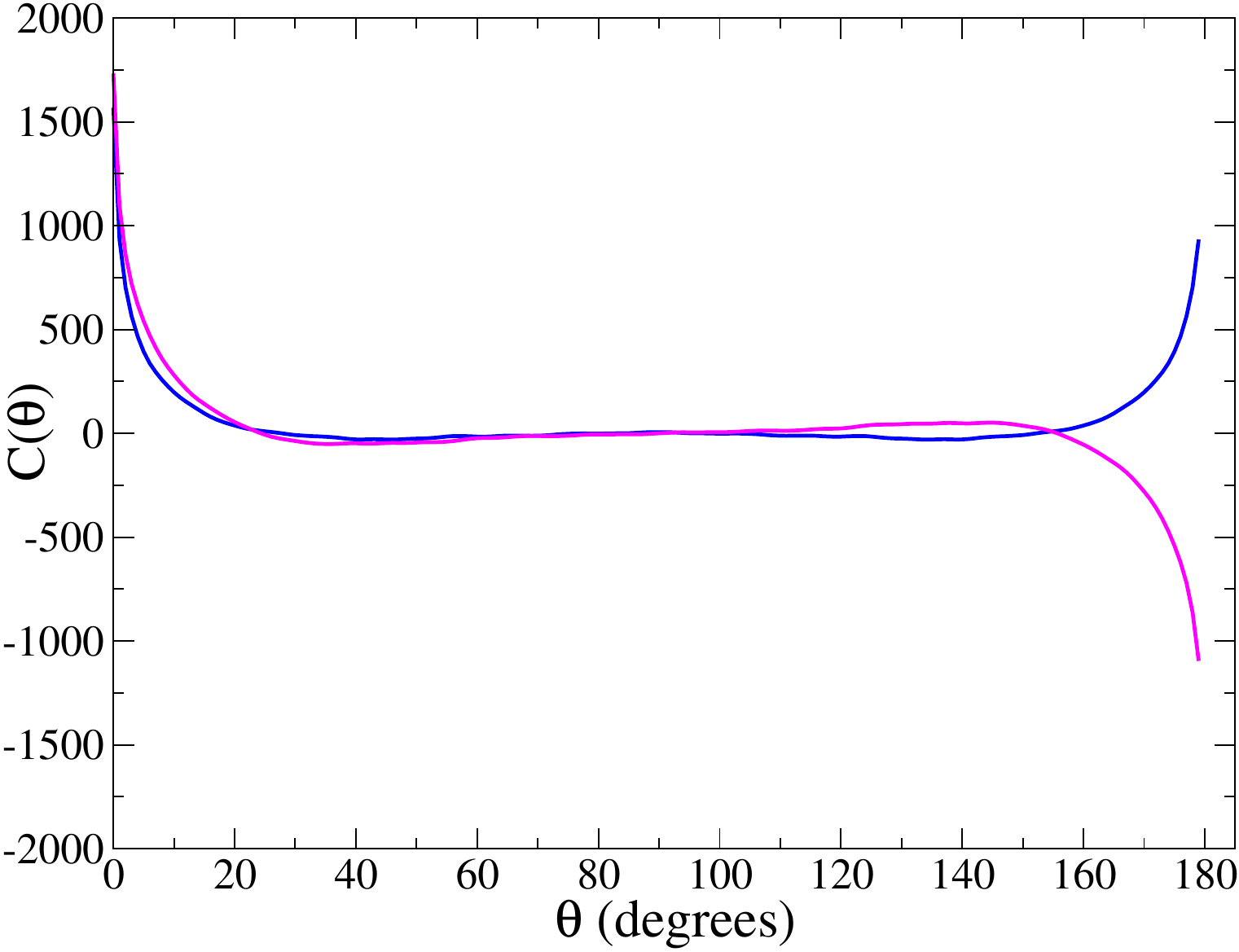}

\caption{\label{fig:2} Even (blue) versus odd (magenta) parity contributions (up to $\ell=200$) to the two-particle correlation function, showing the delicate balance between odd and even polynomials ar large angles, leading either to an upward or a downward tail for even or odd dominance, respectively.}
\end{figure}

\begin{figure}

\centering
\includegraphics[width=8.9cm]{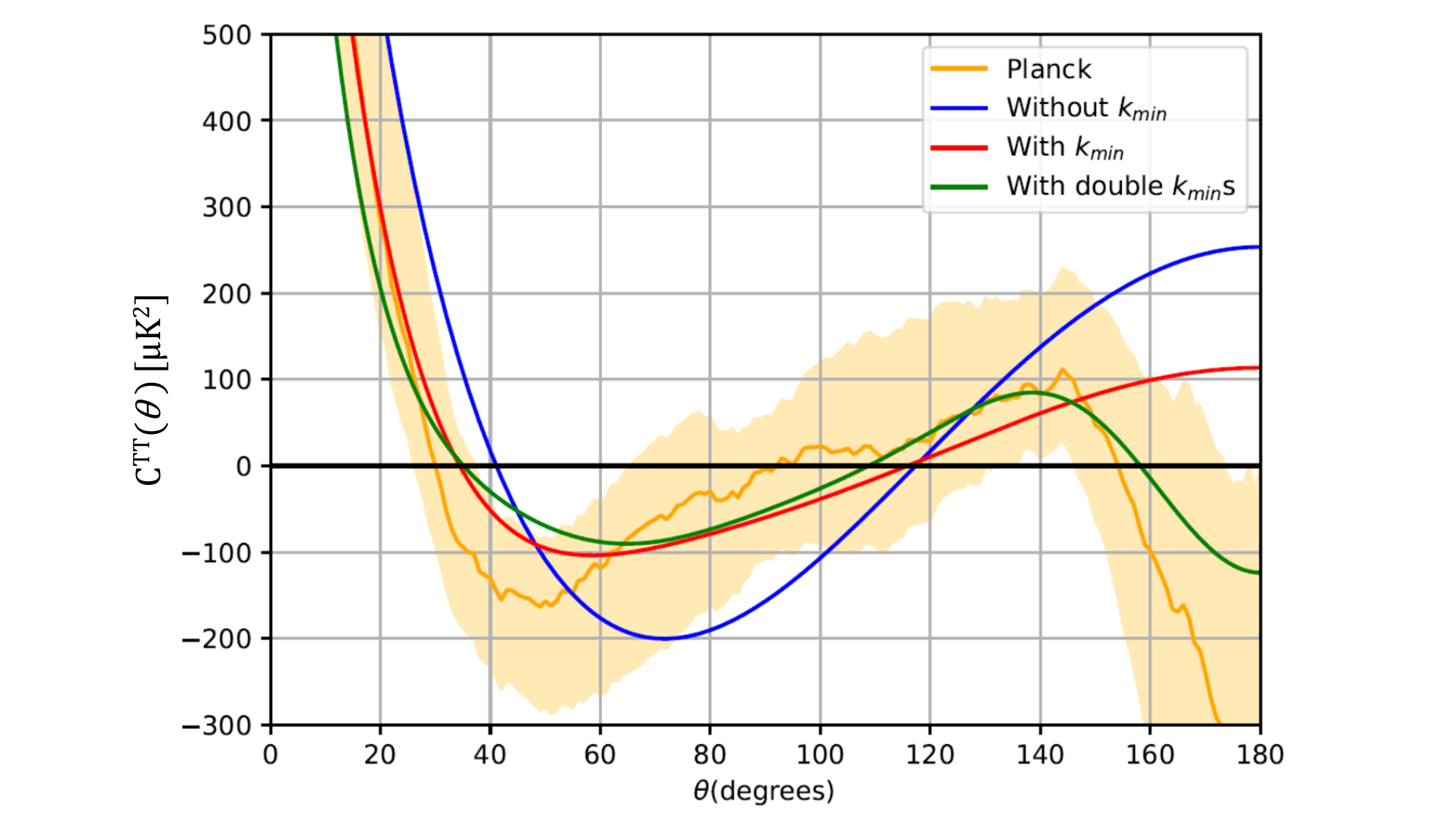}
\includegraphics[width=8.9cm]{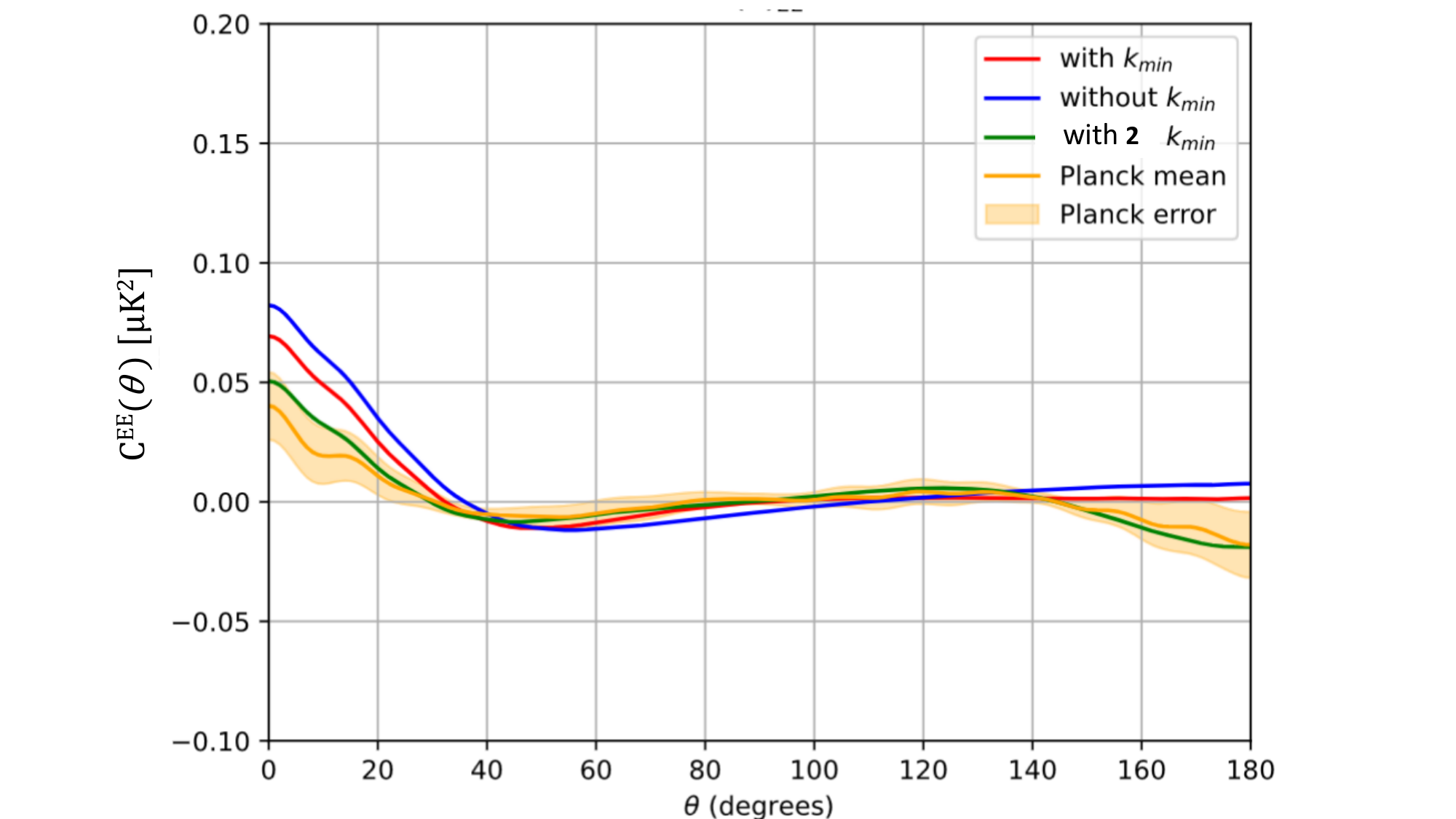}

\caption{\label{fig:3}
Left panel:
Two-point angular temperature correlation function $C^{\rm TT}(\theta)$ with its 1$\sigma$ uncertainty band (orange) from the 2018 {\it Planck} data, together with fits obtained under different assumptions about the IR cutoff(s) in the angular power spectrum.  Blue: no cutoff; Red: single scalar cutoff; Green: scalar cutoff doublet.
Right panel: E-mode polarization two-point correlation function under the same assumptions about the IR cutoff(s) as in the left panel, compared with the {\it Planck} data for $\ell < 30$ with its 1$\sigma$ uncertainty band (orange). Both panels show a downward trend consistent with enhanced odd-parity behavior at low multipoles.}
\end{figure}

In Fig.~\ref{fig:2}, we separately display the odd- and even-parity contributions (from $\ell=2$ up to $\ell=200$) to $C(\theta)$ from scalar modes, computed without imposing any lower cutoff in $C_\ell^{\rm TT}$. A delicate balance between $C_{\rm even}^{\rm TT}(\theta)$ and $C_{\rm odd}^{\rm TT}(\theta)$ is apparent at large angular scales, leading to either positive or negative correlations near antipodal angles. 

Motivated by this behavior, a doublet of IR cutoffs, $k_{\rm min}^{\rm odd/even}$, was introduced instead of a single cutoff as in \cite{Melia:2018,Sanchis-Lozano:2022zrp}, significantly modifying the odd- and even-parity contributions to the angular correlation function.

Using the notation of \cite{Sanchis-Lozano:2022nzp}, we write
\begin{equation}\label{eq:u2}
k_{\rm min}^{\rm odd/even}=
\frac{u_{\rm min}^{\rm odd/even}}{r_L},
\end{equation}
which implements two distinct IR cutoffs in the scalar power spectrum, associated with odd and even multipoles.

Thus, we rewrite the integral of Eq.~\eqref{eq:Cellcutoff} as
\begin{equation}\label{eq:Cellcutoffs}
C_{\ell_{\rm odd/even}}^{\rm TT}{\rm (scalar)} = N_S \int_{u_{\rm min}^{\rm odd/even}\,{\rm (scalar)}}^{\infty} du\, \frac{j_{\ell}^2(u)}{u}\;,
\end{equation}
where the lower limits $u_{\rm min}^{\rm odd/even}$ (see \ref{eq:u2}) affect odd and even multipoles differently, modifying the shape of $C^{\rm TT}(\theta)$.

On the other hand, as already commented, tensor modes affect the temperature correlations in addition to scalar perturbations. The corresponding multipole coefficients $C_{\ell}^{\rm TT}({\rm tensor}),  \ell \ge 2 $, are given by \cite{Mukhanov:2005}
\begin{equation}\label{eq:Celltensoroddeven}
C_{\ell_{\rm odd/even}}^{\rm TT}{\rm (tensor)}\ =\ N_T \ 
\frac{(\ell+2)!}{(\ell-2)!}\ 
\int_{u_{\rm min}^{\rm odd/even}({\rm tensor})}^{\infty}du\ \frac{j_{\ell}^2(u)}{u^5}\;,
\end{equation}
again distinguishing odd and even modes by different
lower cutoffs. Notice the stronger suppression of the integrand due to the fifth power of the variable $u$ in the denominator for tensors, as compared to scalars (see the discussion in Appendix \ref{sec:appB}). This has significant consequences for the computation of the multipole coefficients, since the effect of the cutoff is more pronounced for tensor modes than for scalar ones.

We note that the following hierarchy relation between the lower IR cutoffs of the scalar and tensor sectors was assumed in Ref.\cite{sanchis-sanz:2024}:
\begin{equation}\label{eq:scales}
u_{\rm min}^{\rm even}({\rm scalar}) = 2u_{\rm min}^{\rm odd}({\rm scalar}) =  u_{\rm min}^{\rm odd}({\rm tensor})= u_{\rm min}^{\rm even}({\rm tensor})/2 \; ,
\end{equation}
or equivalently for wavenumbers $k_{\rm min}^{\rm odd/even}$, as $r_L$ is the same for all of them.

Note that the above relation applies to flat geometries and may be modified in warped scenarios. Moreover, we emphasize that, so far, only a single ED has been considered. In Section~\ref{sec:newanal}, toroidal compactification generalizes this relation into a more intricate form (see Appendix A).

If only scalar modes were considered \cite{Sanchis-Lozano:2022nzp}, the following values for the lower cutoffs were obtained: $u_{\rm min}^{\rm odd} = 2.67 \pm 0.31$ and $u_{\rm min}^{\rm even} = 5.34 \pm 0.62$. 
as determined rom a best $\chi^2$ fit of $C^{\rm TT}(\theta)$ to the {\it Planck} temperature anisotropy data.

Upon including tensor modes in the analysis \cite{Sanchis-Lozano:2025csn}, new values for the lower IR cutoffs were obtained from a fit to the same dataset:
\begin{equation}\label{eq:ust}
u_{\rm min}^{\rm odd}({\rm scalar}) = 2.05^{+0.25}_{-0.20} \ ,\quad
u{\rm min}^{\rm even}({\rm scalar}) = 4.10^{+0.25}_{-0.20} ;.
\end{equation}
Because of the strong correlations among the data points, the error bars were estimated through a Monte Carlo procedure, in which 100 mock CMB catalogs were generated and independent fits were performed for each realization (see \cite{Sanchis-Lozano:2022nzp,Sanchis-Lozano:2025csn} for further details).

The respective (mean) comoving wavenumbers are
\begin{equation}
    k_{\rm min}^{\rm odd} \simeq \approx {\rm 1.5} \times 10^{-4}\ {\rm Mpc^{-1}}\ ;\ k_{\rm min}^{\rm even}\approx {\rm 3.0} \times 10^{-4}\ {\rm Mpc^{-1}}\ \to\ k_{\rm min}^{\rm even/odd}\approx
    {\rm few} \times 10^{-42}\ {\rm GeV} \;,
\end{equation}
where we have also expressed the wavenumbers in GeV units. 

The corresponding numerical values for the tensor IR cutoffs are determined from the relation to the scalar ones imposed by our framework, Eq.~\eqref{eq:scales}. We stress that the ratios of the IR cutoffs are fixed by the underlying compactification model and are not treated as free parameters in the fits to the angular correlation data.


\subsection{Temperature angular correlations}\label{sec:Tmode}

Before presenting the analysis performed in this paper, we briefly review the hypotheses introduced in previous studies of temperature angular correlations \cite{Melia:2018}, as well as the corresponding results obtained in \cite{Sanchis-Lozano:2022nzp,Sanchis-Lozano:2025csn}. Both scalar and tensor modes were incorporated through the SW effect, which is expected to dominate at large angular scales. Accordingly, the multipole coefficients in Eq.~\eqref{eq:Csplit} were computed using the standard expressions given in Eqs.~\eqref{eq:Cellcutoffs} and \eqref{eq:Celltensoroddeven}.

{\color{black} First of all, to obtain a realistic representation of the Planck-measured angular correlation function together with its associated $1\sigma$ uncertainty band, use was made of the measured multipole coefficients $C_{\ell}^{\rm TT}$ and their corresponding uncertainties published in \cite{Planck:2018vyg}. To this end, following \cite{Liu:2024mvp}, one thousand mock $C_{\ell}^{\rm TT}$ realizations were generated by sampling the measured $C_{\ell}^{\rm TT}$ values according to their quoted uncertainties, assuming independent half-Gaussian distributions for the upper and lower error bars. For each realization, the corresponding angular correlation function $C(\theta)$ was computed, yielding a sample of one thousand mock $C(\theta)$ curves. This ensemble was then used to estimate the uncertainty band of $C(\theta)$, allowing for asymmetric upper and lower errors through the corresponding half-Gaussian distributions. 

On the other hand, we emphasize that, within the large-angle approximation adopted in this work, the angular two-point temperature correlation function depends on a remarkably small number of effective parameters. Assuming that the SW contribution dominates the temperature anisotropies and adopting a scale-invariant primordial spectrum ($n_s=1$), the correlation function is determined essentially by an overall normalization and the dimensionless integration variable $u$. If primordial tensor perturbations are included, an additional dependence on the tensor-to-scalar ratio $r$ arises through their contribution to the low-multipole temperature anisotropies. By contrast, the remaining cosmological parameters of the standard $\Lambda$CDM model enter only weakly within these approximations, their effects being largely absorbed into the overall normalization. 

This simplification is, however, specific to the two-point correlation function in the SW regime. Higher-order correlation functions, such as the bispectrum and trispectrum, are generally sensitive to nonlinear evolution, mode coupling, and primordial non-Gaussianity, thereby introducing explicit dependence on a much broader set of cosmological parameters.}

Fig.~\ref{fig:3} shows the two-point correlation function $C^{\rm TT}(\theta)$, fitted to {\em Planck} 2018 data (orange), over the angular range $4^{\circ}$–$180^{\circ}$ under different assumptions for the infrared cutoff(s): (i) no cutoff, (ii) a single scalar IR cutoff, and (iii) doublets of scalar and tensor IR cutoffs. This figure is reproduced from \cite{Sanchis-Lozano:2025csn}. 

The main features of the {\em Planck} data, including the change in the sign of the slope of the fitted curve at about $140^{\circ}$, leading to a downward tail with negative values, are well reproduced in the doublet case. Additional contributions from tensor modes only marginally improve the fit, as their impact is largely confined to large angular scales. 

\subsubsection*{Physical scale of the infrared cutoffs}

Estimating the energy scale associated with the IR cutoffs in our framework and verifying its consistency with cosmic evolution is crucial, as it indicates the epoch at which compactification occurred. Physical cutoffs are obtained by converting comoving quantities into physical ones, namely by dividing them by the cosmic scale factor. In particular, this conversion must be performed at the time when compactification of the extra dimensions (ED) took place, denoted by $t_{\rm extra}$. The corresponding physical wavenumber (essentially set by the inverse radius of the ED) is therefore given by $k_{\rm min}/a(t_{\rm extra})$. 

Assuming that compactification happened at some time between the end of the Planck epoch $t_{\rm Planck}$ and the beginning of
inflation $t_{\rm init}$, the respective scale factors 
$a(t_{\rm Planck})\simeq 10^{-61}$ and $a(t_{\rm init})\simeq 10^{-56}$ \cite{Liu:2024zql} lead to the
following physical cutoff range:
\begin{equation}\label{eq:kscale}
    \frac{k_{\rm min}}{a(t_{\rm extra})} \in 10^{19}-10^{14}\ {\rm GeV},\ \ t_{\rm Planck} < t_{\rm extra} < t_{\rm init} \;,
\end{equation}
which contains the GUT scale. Thus, the physical scale of the infrared cutoff(s) in the CMB temperature correlations is consistent with the assumption of an ED with a high compactification scale, above the Planck epoch, thereby avoiding the trans-Planckian problem. 

For later reference, it is useful to quote the following expression relating the energy scale of inflation to the tensor-to-scalar ratio $r$:
\begin{equation}\label{eq:energyscale}
V^{1/4} \approx 1.04 \times 10^{16}\ \mathrm{GeV} \times (r/0.01)^{1/4}.
\end{equation}
in single-field slow-roll models. A significantly large tensor amplitude would provide strong evidence that inflation took place at an extremely high energy scale, such as the Grand Unified Theory (GUT) scale, as we will later see.

Next, before embarking on the new analysis of Sec.~\ref{sec:newanal}, we revisit E-mode polarization correlations as an additional test of the existence of IR cutoffs in the primordial spectrum.

\subsection{E-mode polarization correlations}\label{sec:Emode}

Let us recall that E-mode polarization in the CMB arises primarily from Thomson scattering of photons off free electrons in the presence of quadrupole anisotropies. The dominant contribution comes from density (scalar) perturbations at recombination, while re-ionization acts as a second round of Thomson scattering that generates additional E-mode polarization, especially on large angular scales. Gravitational lensing also contributes to the overall E-mode signal. 
 
 As argued in \cite{Sanchis-Lozano:2025csn}, E-mode polarization in the CMB can also provide valuable insights into the very early Universe, as it is sensitive to a possible parity multipole imbalance particularly at low  $\ell$ values. 

The angular two-point correlation function for the E-mode polarization can be written as \cite{Liu:2025yvp}
\begin{equation}\label{eq:CEE}
C^{\rm EE}(\theta)=\sum_{\ell \ge 2} \frac{(2\ell+1)}{4\pi}\frac{(\ell+2)!}{(\ell-2)!}\ C_{\ell}^{\rm EE} \ P_{\ell}(\cos{\theta}), .
\end{equation}
The $(\ell+2)!/(\ell-2)!$ factor has been explicitly factored out from the coefficient definition to highlight that polarization correlations are strongly dominated by large-$\ell$ multipoles, thereby diminishing the impact of any IR cutoff. Nonetheless, useful information associated with large angular scales, i.e., low $\ell$, can still be obtained from Eq.~\eqref{eq:CEE}.

In the right panel of Fig.~\ref{fig:3} we reproduce the corresponding curve for the E-mode polarization correlation function from \cite{Sanchis-Lozano:2025csn} for $\ell<30$, using the same parameter values obtained from the fit to the CMB power spectrum. A downward tail again emerges at large angular scales, reflecting the odd-parity preference observed in the temperature correlations. Although this plot does not result from a new fit, the behavior of the curves provides an important cross-check between two independent observables, thereby supporting the need for an IR cutoff doublet.

\section{New analysis of angular correlations}\label{sec:newanal}

In this work, we reexamine large angular scales detail. The main differences 
with respect to previous work reported in the precedent section, showing the main lines of our new approach, are  
\begin{itemize}
\item Rather than performing a fit to the CMB temperature correlation function, we take a fit to temperature angular power spectrum as our starting point. Comparisons with angular correlations will be considered as a consistency check. 
\item  When generalizing from a single extra dimension to a compactification on three tori, the numerical implementation incorporates six tensor contributions while retaining a single effective scalar field. No additional scalar contributions are considered. 



We will include contributions arising from EDs according to the theoretical framework developed in section \ref{sec:torcomp}.

\end{itemize}

Furthermore, a smoothing procedure, explained in more detail in Appendix B, is applied instead of hard cutoffs
\begin{equation}\label{eq:Cellcutoffoesm}
C_{\ell_{\rm odd/even}}{\rm (scalar)}\ =\ N_S\ \int_0^{\infty}\ du\ f(u; u_{\rm min}^{\rm odd/even},\Delta_u)\ \frac{j_{\ell}^2(u)}{u}\;,
\end{equation}
where we have assumed $\Delta_u/u_{\rm min} = 0.1$ for both even and odd hard IR cutoffs. This assignment follows from the uncertainty in the extraction of the numerical values for 
the lower cutoffs in Eq.\ref{eq:ust}. We have explicitly checked that by varying the above ratio by a factor of two does not yield significant deviations in our later development.

We assign the value of $\Delta_u$ to the uncertainty in the numerical determination of $u_{\rm min}$ obtained from the fit of {\it Planck} data to the two-point correlation function, as described in \cite{Sanchis-Lozano:2022nzp}. Later we distinguish between odd and even modes by a different $\Delta_u^{\rm odd/even}$ width in the smoothing function \ref{eq:fsmooth}.

In practice, the impact of the above described smoothing procedure becomes relevant only at rather low multipoles, i.e., $\ell \lesssim 4$, see appendix \ref{sec:appB}. Actually, the odd-parity dominance \cite{Land:2005jq,Copi:2010na,Creswell:2021eqi} could not be fully achieved by using a single cutoff, whose main effect on the multipole coefficients is limited to the first few terms, especially the quadrupole.

\begin{figure}

\centering
\includegraphics[width=8.9cm]{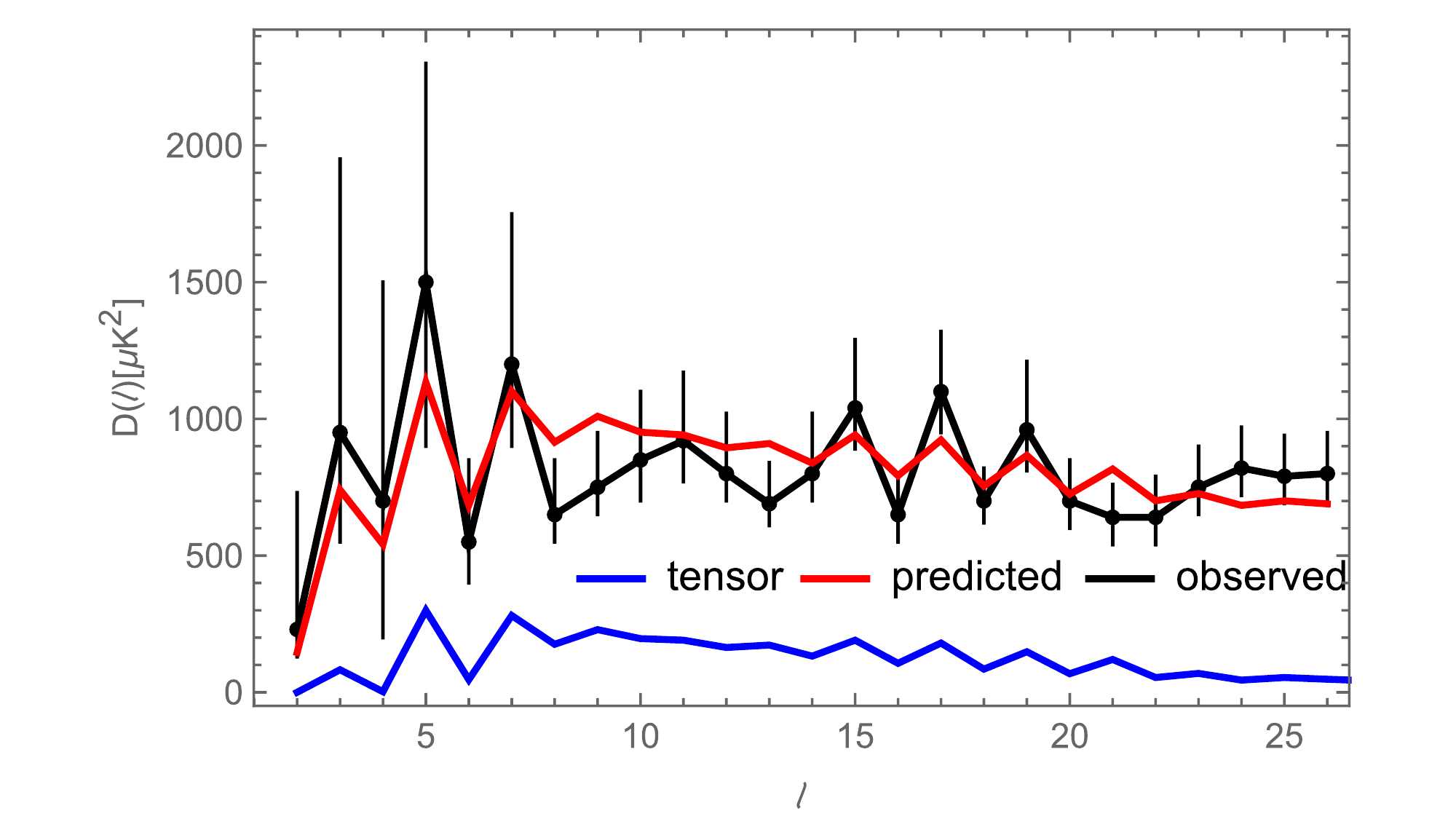}
\includegraphics[width=8.9cm]{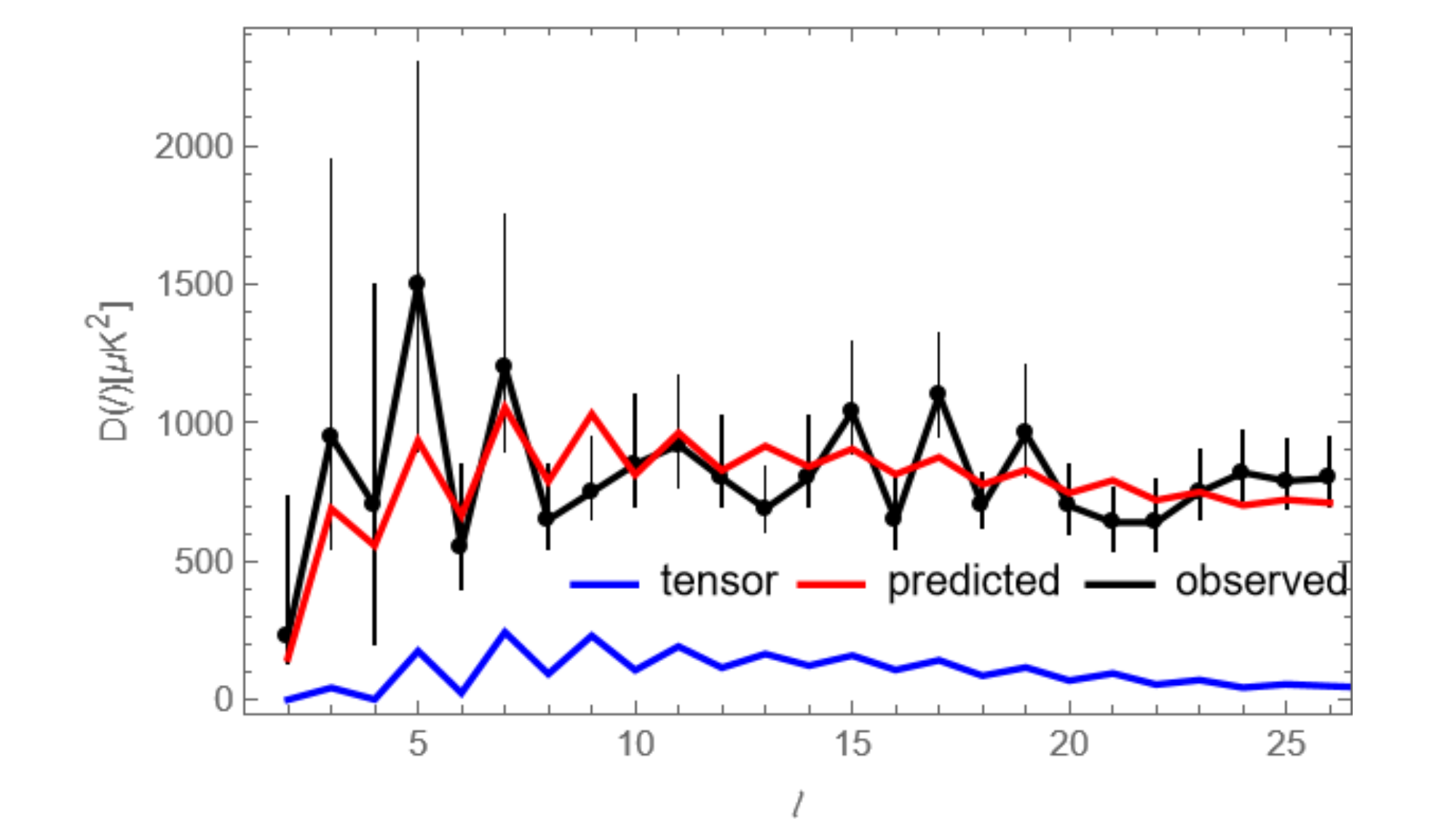}
\caption{\label{fig:4} Temperature angular power spectrum $D_{\ell}$ for $\ell < 30$, showing the experimental data and the $C_{\ell}$ values obtained from a best-fit $\chi^2/{\rm d.o.f.}$ analysis. The contribution from tensor modes to the multipole coefficients is represented by the lower blue curve. Left panel: $\gamma = 1$; right panel: $\gamma = \sqrt{3}$. In both panels, the observed and theoretical points are connected by straight segments to better visualize the postulated underlying sawtooth structure, characterized by successive peaks and valleys, as a consequence of the IR cutoffs.}
\end{figure}

\begin{figure}

\centering
\includegraphics[width=8.9cm]{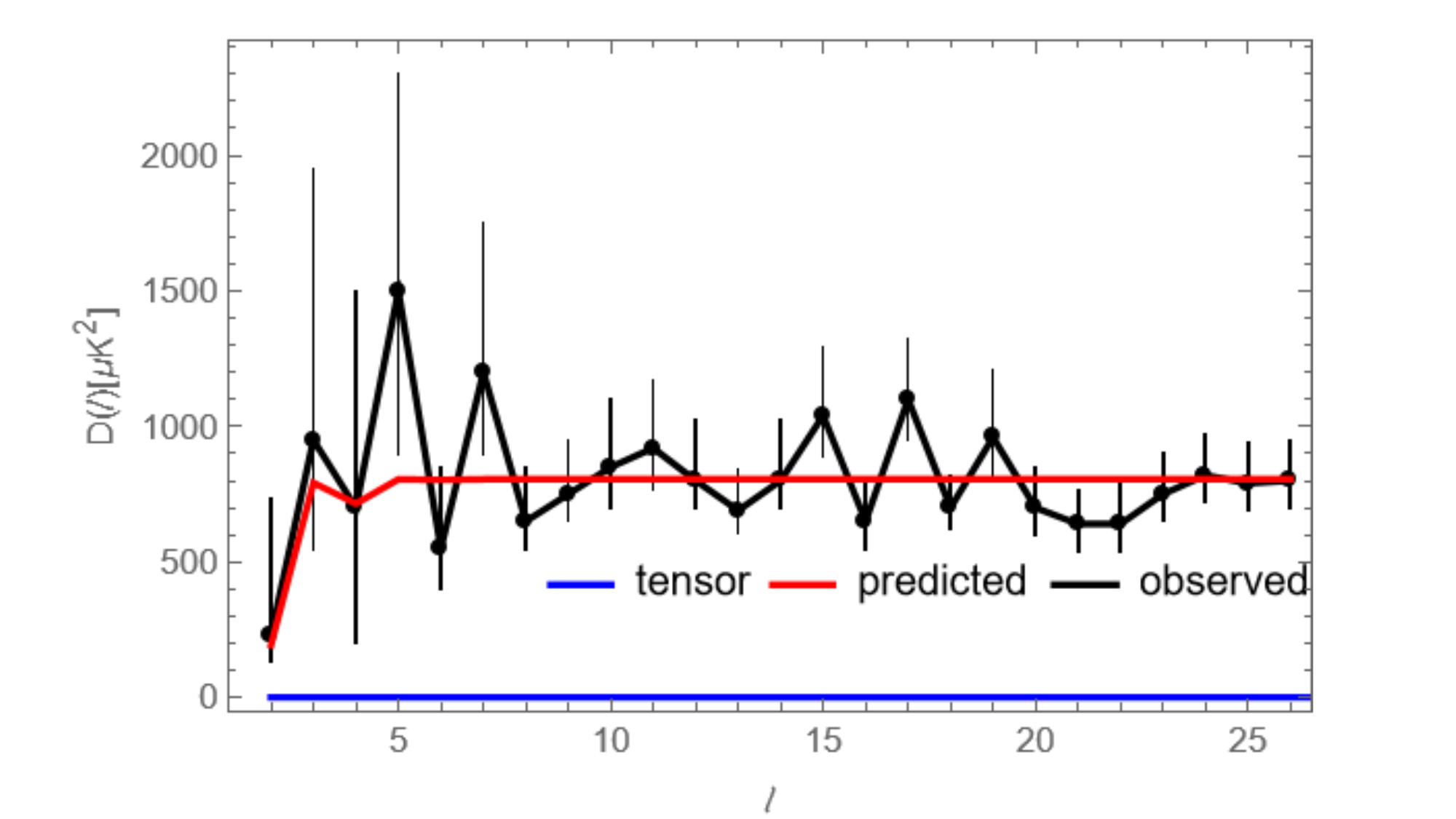}
\includegraphics[width=8.9cm]{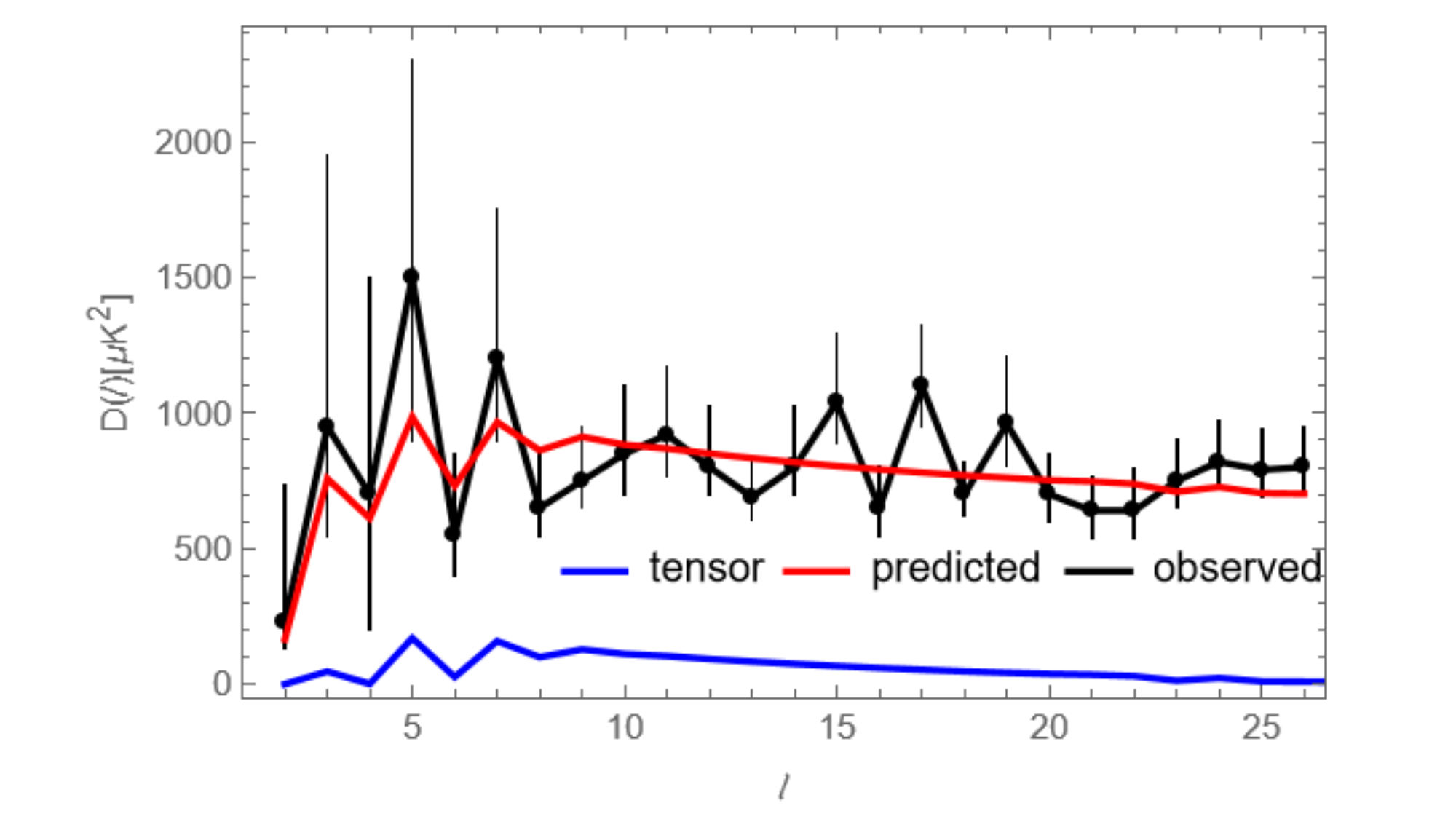}
\caption{\label{fig:5} The same as in Fig.\ref{fig:4}. Left panel: only scalar modes (no tensor modes) with an IR cutoff doublet contributing to the predicted curve (in red); Right panel: both scalar modes and tensor modes from a single ED and two IR cutoff doublets contributing to the predicted curve (in red).}
\end{figure}

For the tensor modes, we employ a similar smoothing procedure as for scalar modes according to
\begin{equation}\label{eq:Cellcutoffotsm}
C_{\ell_{\rm odd/even}}{\rm (tensor)}\ =\ N_T \ 
\frac{(\ell+2)!}{(\ell-2)!}\ 
\int_0^{\infty}du\ f(u; u_{\rm min}^{\rm odd/even},\Delta_u)\ \frac{j_{\ell}^2(u)}{u^5}\;,
\end{equation}
Assuming that the physical effect behind the smoothing of the hard cutoffs is the same for scalar and tensor modes, we adopt the same smoothing function as in Eq.~\eqref{eq:Cellcutoffoesm}.

As explained in the Introduction and Section~\ref{sec:scaltens}, we assume a toroidal geometry such that, upon compactification, additional EDs emerge, potentially influencing present-day CMB observations. In typical string-theory constructions, six extra spatial dimensions arise; together with the three extended spatial dimensions and one time dimension, this leads to a ten-dimensional spacetime, nine of whose dimensions are spatial. 

 At this point, let us recapitulate the main theoretical inputs involved in our calculations. For each $i$-torus ($i=1,2,3$), two extra dimensions (EDs) emerge, each characterized by a radius $R_{i,I}$ ($I=1,2$), whose inverse sets the energy scale for the corresponding IR cutoff. In turn, each IR cutoff splits into an IR doublet associated with the odd or even parity of the multipoles contributing to the scalar and tensor power spectra of the CMB. It is crucial to emphasize that these IR cutoffs are not free parameters in our analysis, but are instead related to one another through the model conditions summarized in Appendix \ref{sec:appA}.

In Fig.\ref{fig:4}, we plot the observed angular power spectrum for $\ell < 30$ alongside the theoretical expectations obtained from a fit including both scalar and tensor contributions within our framework, for $\gamma = 1$ (left panel) and $\gamma = \sqrt{3}$ (right panel). In both panels we adopt the existence of 3 tori leading to 6 compactified EDs and the corresponding IR cutoffs as explained before. To better visualize the underlying pattern of successive peaks and valleys, we connect both the data and predicted  points with straight segments, highlighting the sawtooth structure of the power spectrum generated by both scalar and tensor modes as a consequence of the difference between odd and even cutoffs. {\color{black} Remarkably, a similar pattern arises in the scalar power spectrum in the alternative framework proposed in \cite{Gaztanaga:2024vtr} and further developed in \cite{Gazta_aga_2025,Gaztanaga:2025awe}, based on a completely different model invoking a new view of quantum physics.}

On the other hand, it is worth emphasizing that, {\color{black} in our approach,} scalar modes mainly affect low-$\ell$ multipoles (especially the quadrupole, $\ell = 2$), whereas tensor modes introduce additional contributions (each with their corresponding IR cutoffs) arising from the EDs postulated in this work. This behavior is a mathematical consequence of the larger values of the tensor IR cutoffs compared to the scalar ones, which leads to a stronger impact on higher multipoles (see Appendix~\ref{sec:appA}). To better illustrate this key point, we show in Fig.~\ref{fig:5} similar plots to those of Fig.~\ref{fig:4}, now: (i) Left panel: No tensor modes and an IR cutoff doublet from a single ED; Right panel: both scalar and tensor modes with their respective IR cutoff doublets for a single ED, in agreement with the analysis presented in \cite{sanchis-sanz:2024}.

According to our best $\chi^2$ fit to the temperature angular power spectrum restricted to low multipoles ($\ell < 30$), as shown in Fig.~\ref{fig:4}, we obtain
\begin{equation}\label{eq:ustnew}
u_{\rm min}^{\rm odd}(\mathrm{scalar}) = 2.00^{+0.25}_{-0.20} \ ,\quad
u_{\rm min}^{\rm even}(\mathrm{scalar}) = 4.00^{+0.25}_{-0.20}\; ,
\end{equation}
The respective tensor IR cutoffs can be readily obtained from these values under the model assumptions adopted in this work.

The lower limits $u_{\rm min}^{\mathrm{odd/even}}$ shown in Eq.~\ref{eq:ustnew} are slightly smaller than those reported in Eq.~\eqref{eq:ust}, although fully compatible within uncertainties. The decrease can be qualitatively understood as follows: as more tensor contributions are included, each with their own IR cutoffs, the scalar contribution is less required to break the parity balance in the angular correlations.

We obtain $\chi^2_{\rm d.o.f.} = 0.60/0.76$ for the choices $\gamma = 1$ and $\gamma = \sqrt{3}$, respectively, where $\gamma$ is defined in Eq.~\eqref{ratioR}. The corresponding p-value turns out to be approximately 0.9.

Admittedly, the fit may appear “too good”, likely as a consequence of the large error bars in the power spectrum, which are moreover highly correlated. A similar conclusion is reached from the fit to the correlation function $C^{\rm TT}(\theta)$ over the full angular range ($4^\circ$–$180^\circ$), shown in the left panel of Fig.~\ref{fig:3}. The values obtained for $u_{\rm min}^{\rm odd/even}$ (for both scalar and tensor modes), and error bars, turn out to be essentially the same.

In the next section, we investigate parity imbalance as a means to enhance our sensitivity to different early-Universe topologies, the main goal of this paper.
\begin{figure}

\centering
\includegraphics[width=8.5cm]{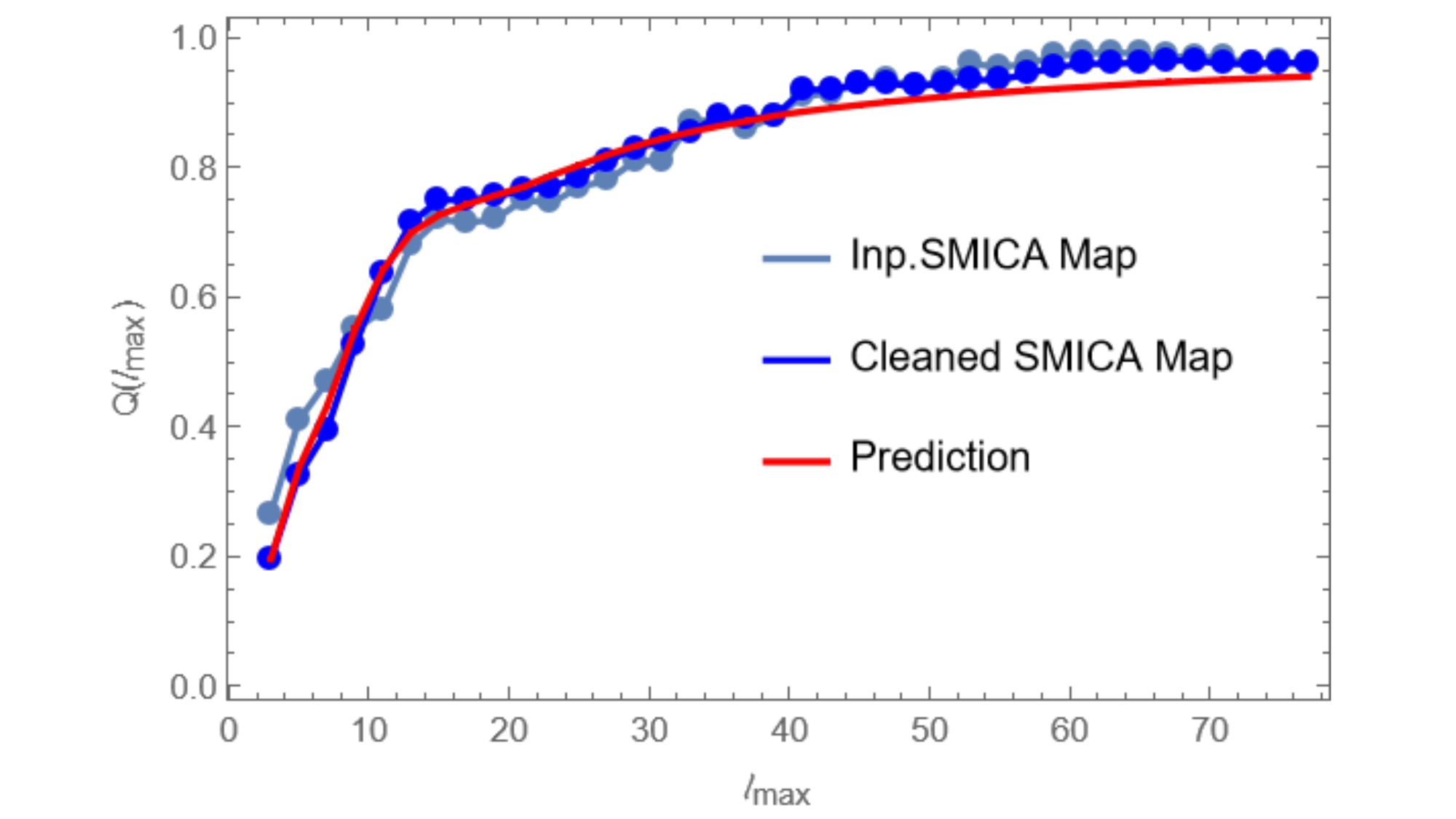}
\includegraphics[width=8.5cm]{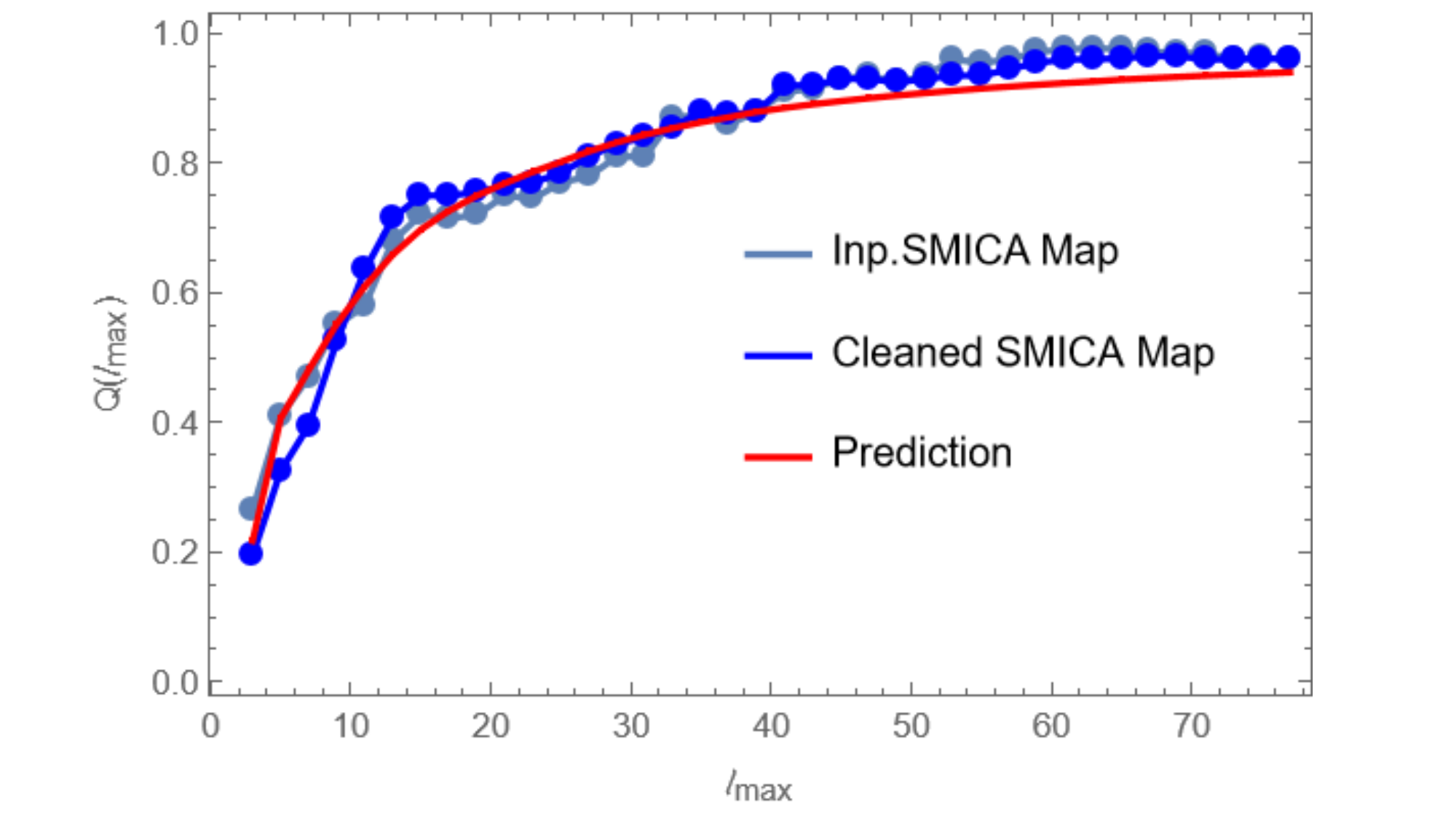}

\centering
\includegraphics[width=8.5cm]{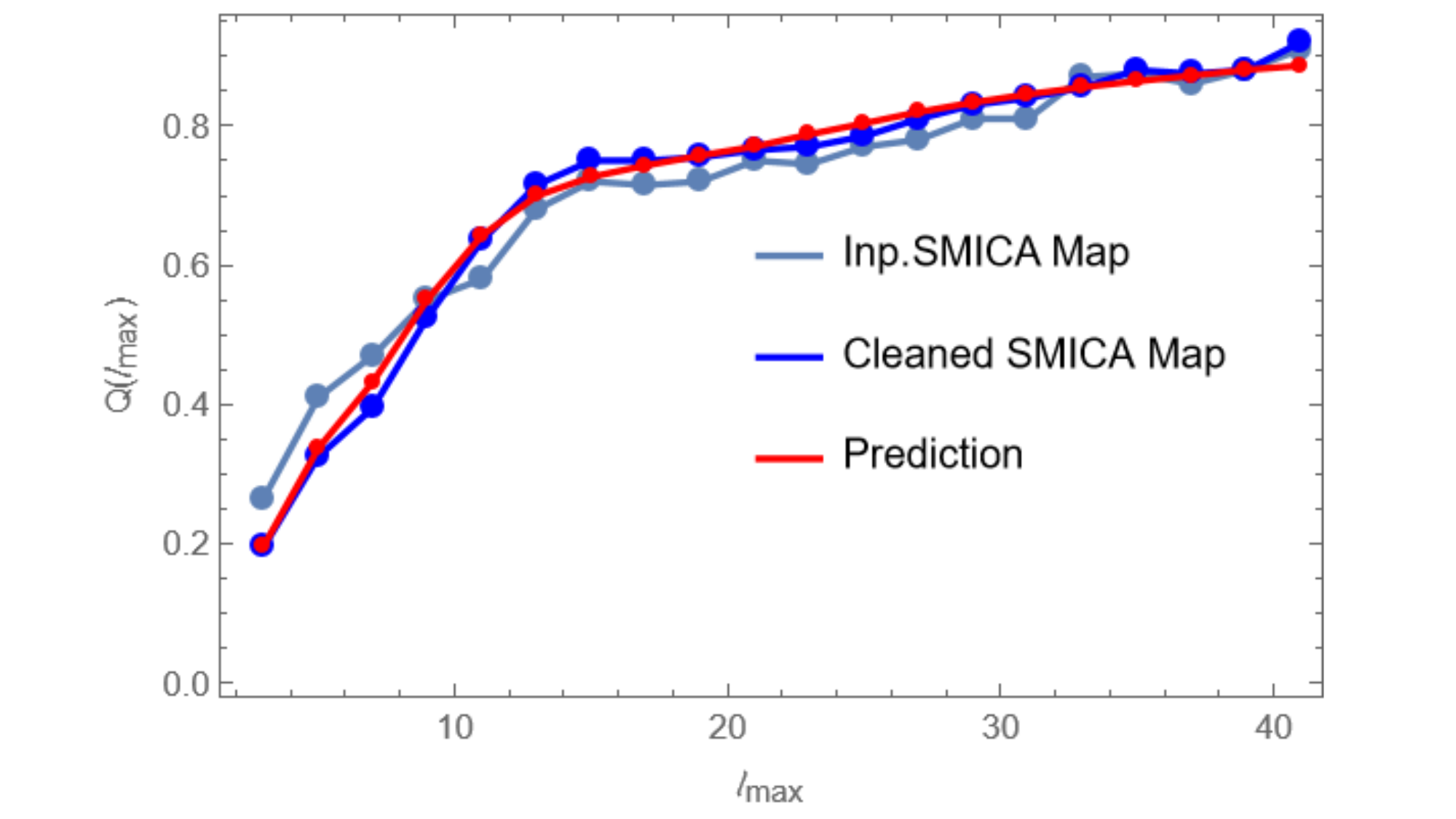}
\includegraphics[width=8.5cm]{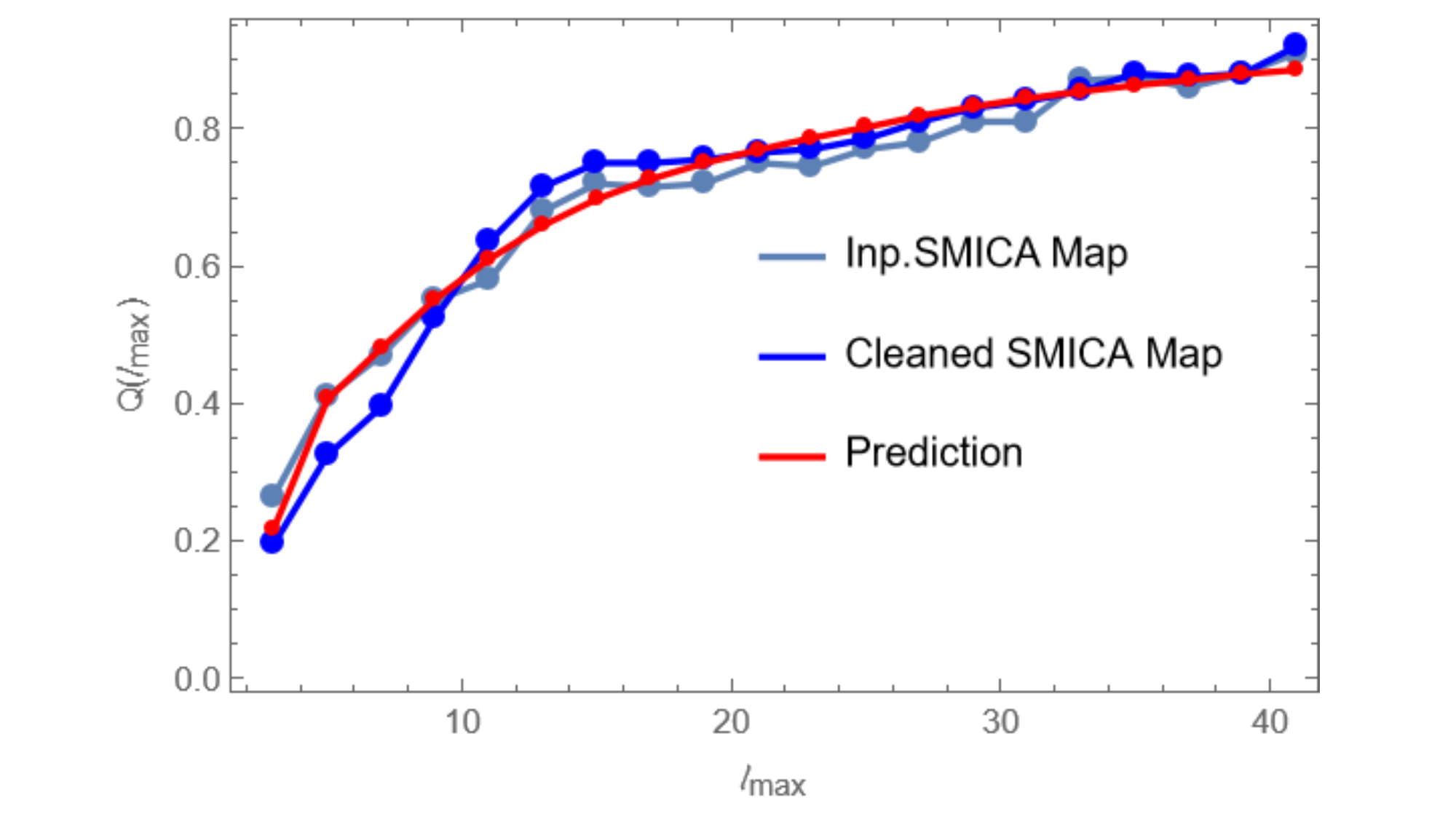}

\caption{\label{fig:6} Upper panels: Parity statistic $Q^{\rm TT}(\ell_{\rm max})$ versus $\ell_{\rm max}$ for different SMICA masks \cite{Panda2021}, versus the theoretical prediction (in red) for $\ell < 77$. Left panel: $\gamma=1$; Right panel: $\gamma=\sqrt{3}$.
Lower panels: the same for $\ell < 41$. {\color{black} The "predicted" curve (in red) corresponds to both scalar and tensor modes}.}
\end{figure}

\begin{figure}

\centering
\includegraphics[width=8.5cm]{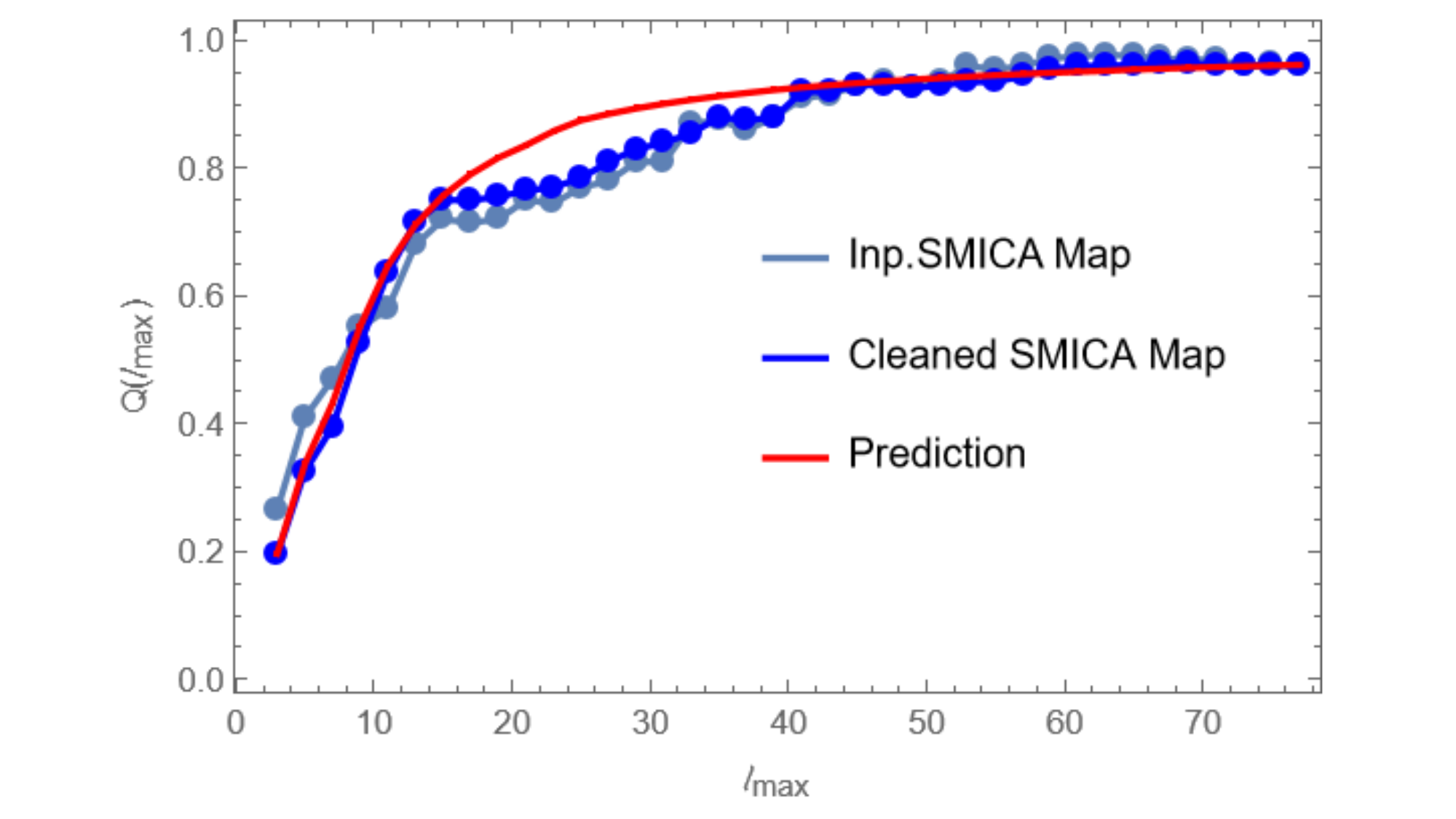}
\includegraphics[width=8.5cm]{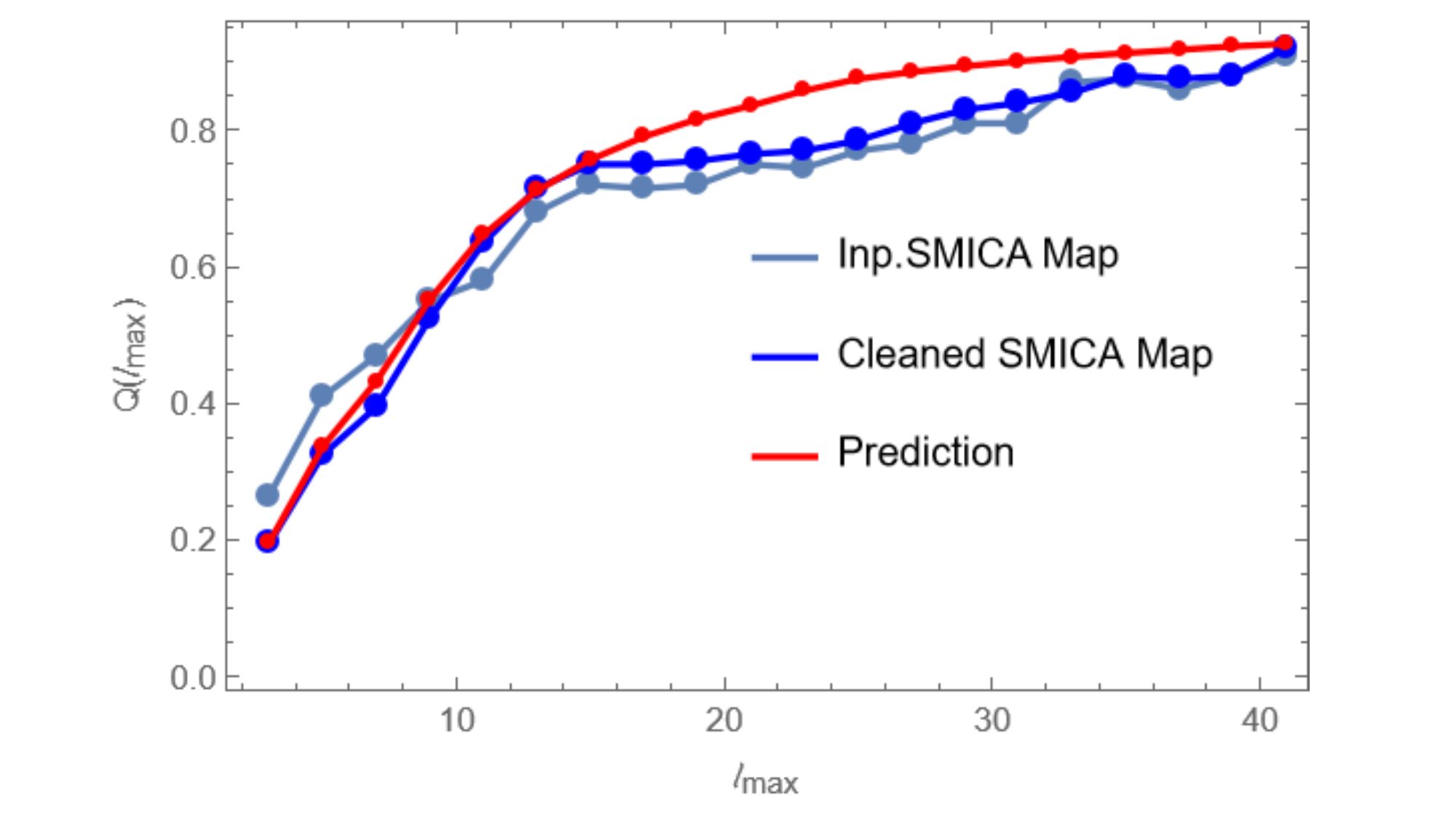}

\caption{\label{fig:7}  Parity statistic $Q^{\rm TT}(\ell_{\rm max})$ versus $\ell_{\rm max}$ for different SMICA masks \cite{Panda2021}, versus the theoretical prediction for $\gamma = 1$. Left panel: $\ell < 77$ and right panel $\ell < 41$. {\color{black} The "predicted" curve (in red) corresponds to scalar modes only.}}
\end{figure}

\section{Parity analysis}

The deviation from even–odd parity balance in CMB angular correlations can be quantified using the parity statistic: \cite{Aluri2012,Panda2021} 
\begin{equation}\label{eq:Qstat}
Q^{\rm TT}(\ell_{\rm max})=\frac{2}{\ell_{\rm max}^{\rm odd}-1}\sum_{\ell=3}^{\ell_{\rm max}^{\rm odd}}\ 
 \frac{D_{\ell-1}}{D_{\ell}}\,,  \qquad\ell_{max}^{\rm odd} \ge 3\;,
\end{equation}
where $\ell_{\rm max}^{\rm odd}$ denotes the maximum odd multipole considered, and $D_{\ell}$ the CMB temperature angular power spectrum.

This statistic compares the power in consecutive even and odd multipoles: if the CMB fluctuations were perfectly parity symmetric, even and odd multipoles would contribute equally and $Q \approx 1$. Values $Q>1$ indicate a relative enhancement of power in the even multipoles, while $Q<1$ signals a preference for odd multipoles. In this way, $Q^{\rm TT}(\ell_{\rm max})$ provides a simple, quantitative measure of parity asymmetry over a range of angular scales, smoothing the abrupt oscillations shown when using other related statistics 
\cite{Sanchis-Lozano:2022zrp}.

Notice that, although $Q^{\rm TT}(\ell_{\max})$ is defined as a ratio of multipole coefficients, such that the overall normalization cancels, it retains a residual but still significant dependence on the relative scalar and tensor contributions, (and hence on their ratio $r$).

We emphasize that the following statistical analysis of $Q^{\rm TT}(\ell_{\max})$ should be interpreted as a conditional goodness-of-fit test of the power spectrum shown in Fig.~\ref{fig:4}, rather than as a new independent fit, since it relies on parameter values previously determined in our analysis of the power spectrum.

In Fig.~\ref{fig:6} we present a set of plots of $Q^{\rm TT}(\ell_{\rm max})$ as a function of $\ell_{\rm max}$, having set $r=0.02$, providing a “zoomed-in” view of the low-$\ell$ region, or equivalently, large angular scales, where effects from EDs in the early Universe are mainly expected to manifest. The upper panels show the multipole range up to $\ell_{\rm max}=77$ for $\gamma=1$ (left) and $\gamma=\sqrt{3}$ (right). The lower panels display an even more restricted range, $\ell_{\rm max}<41$, where the impact of tensor modes on temperature anisotropies is expected to be more significant.

\begin{figure}
\centering
\includegraphics[width=8.5cm]{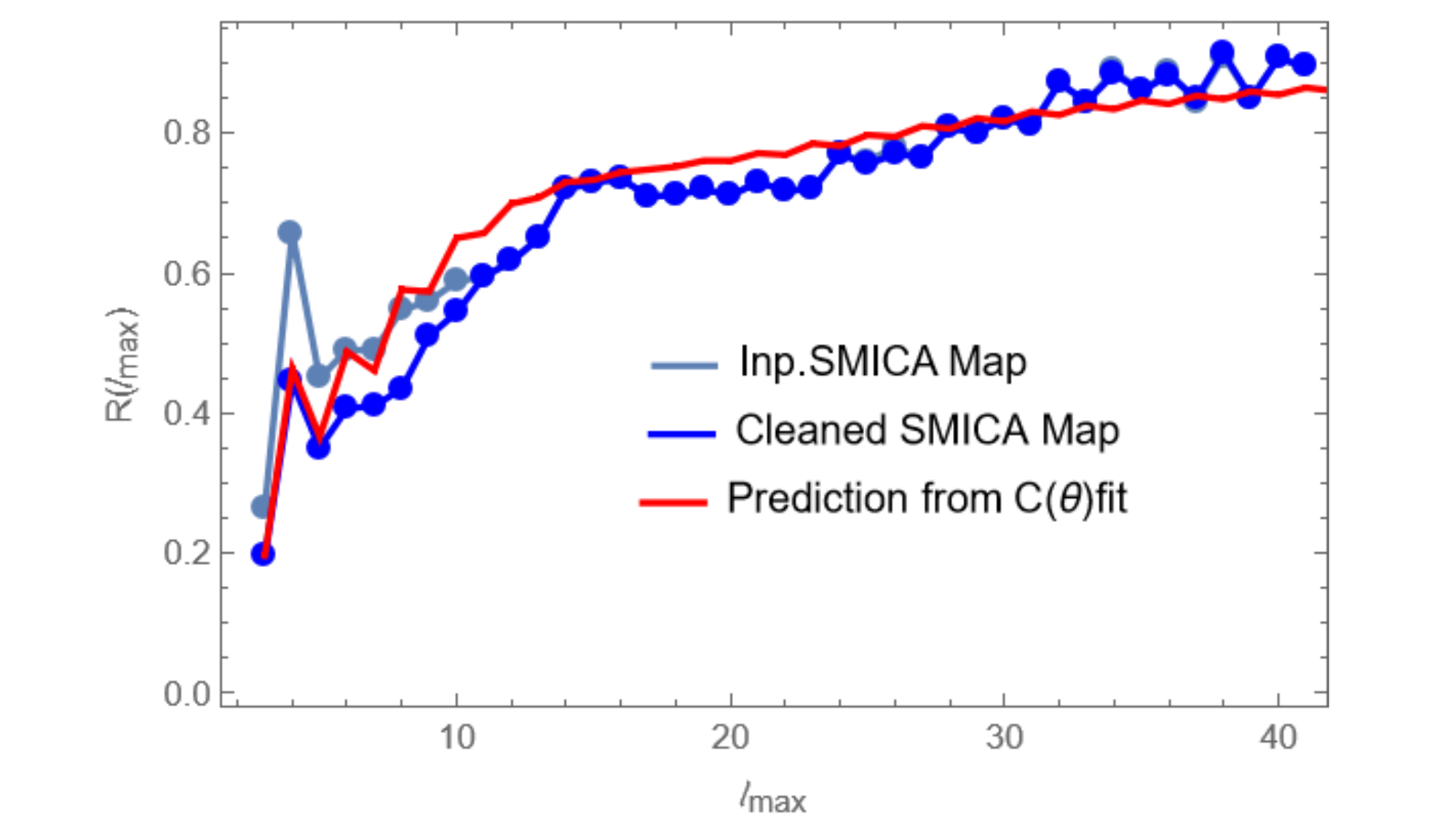}
\includegraphics[width=8.5cm]{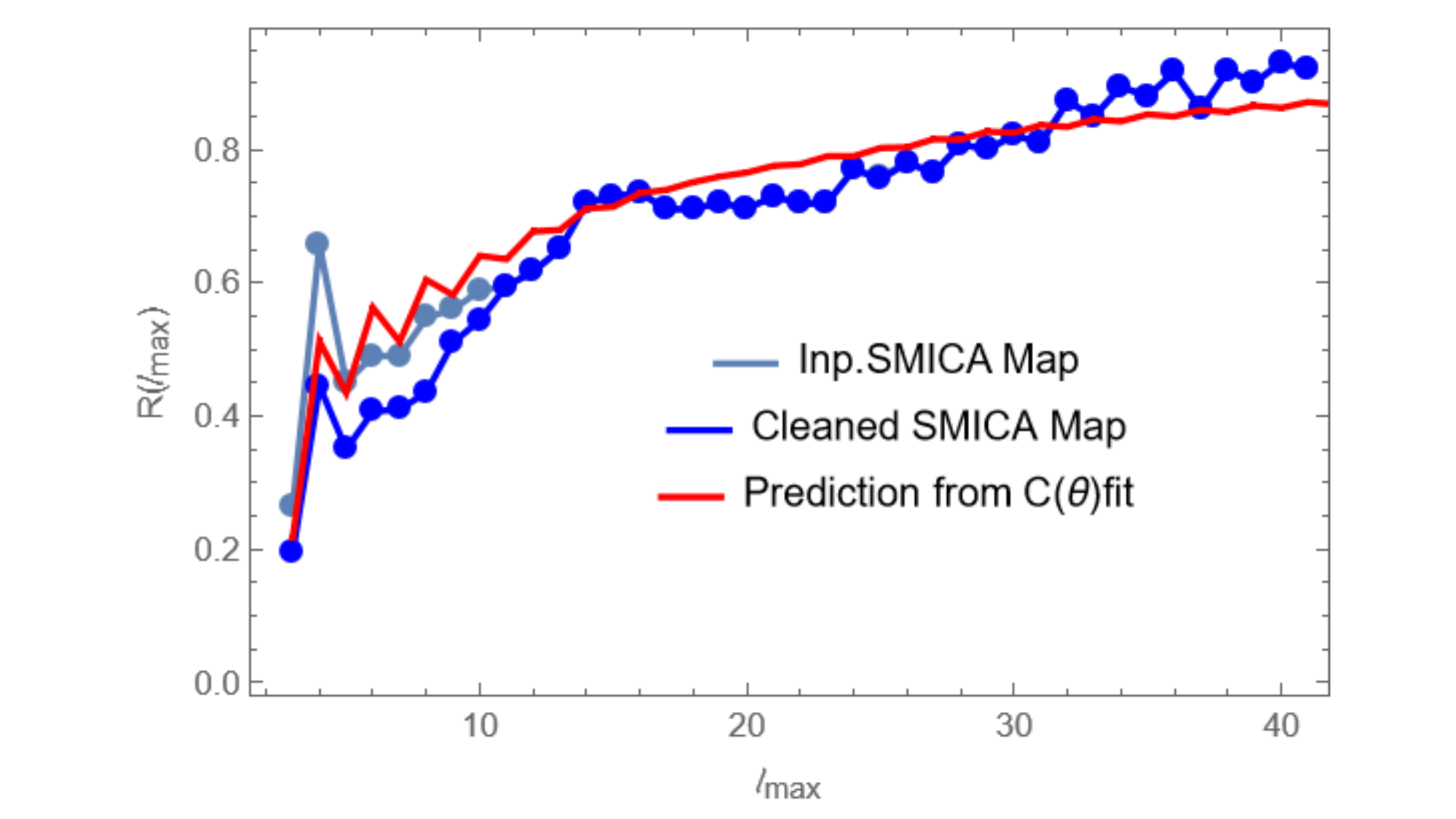}
\caption{{\color{black} $R^{TT}$ statistic as a function of $\ell_{\rm max}$ for $\gamma=1$ (Left panel) and $\gamma = \sqrt{3}$ (Right panel), showing odd-parity dominance.}}\label{RTT-2}
\end{figure}

In all panels of Fig.~\ref{fig:6}, the angular data exhibit a preference for odd-parity multipoles, consistent with the behavior of the temperature correlation function $C^{\rm TT}(\theta)$ observed previously. The origin of this long-standing imbalance has been attributed in the literature either to an unknown physical mechanism or to underestimated uncertainties, including statistical fluctuations. In this work, we instead propose a physical explanation based on the influence of EDs arising from a toroidal compactification of the early Universe.

{\color{black} Alternatively, the parity asymmetry of the CMB temperature anisotropies is commonly
quantified by the statistic introduced in the Planck analyses,
\begin{equation}
R^{TT}(\ell_{\max})
\equiv
\frac{D^{TT}_{+}(\ell_{\max})}
     {D^{TT}_{-}(\ell_{\max})},
\label{eq:RTT_Planck}
\end{equation}
where
\begin{equation}
D^{TT}_{\pm}(\ell_{\max})
=
\frac{1}{\ell_{\rm tot}^{\pm}}
\sum_{\substack{\ell=2\\ \ell\,\mathrm{even/odd}}}^{\ell_{\max}}
\frac{\ell(\ell+1)}{2\pi}\,C_\ell^{TT}.
\label{eq:Dpm}
\end{equation}
Here, the $+$ ($-$) subscript denotes the sum over even (odd)
multipoles, and $\ell_{\rm tot}^{\pm}$ is the number of multipoles
included in the corresponding sum. Thus, $R^{TT}<1$ indicates a
relative excess of power in the odd multipoles, whereas
$R^{TT}>1$ indicates a relative excess in the even multipoles.}

 Let us stress that tensor modes play a fundamental role in the analysis, as they induce a relative suppression of $Q^{\rm TT}(\ell_{\max})$ ({\color{black} or equivalently $R^{\rm TT}$)} over the multipole range $10 \lesssim \ell_{\max} \lesssim 40$ (see Fig.~\ref{fig:6}), given the IR cutoff values adopted in our framework, which extend the effect to higher multipoles. {\color{black} According to our approach,} no comparable effect can be achieved by including scalar modes alone. 

To further clarify this point, Fig.~\ref{fig:7} displays the same curve of $Q^{\rm TT}(\ell_{\max})$ as a function of $\ell_{\max}$, but computed using exclusively scalar modes derived from the fit to the power spectrum shown in Fig.~\ref{fig:5}. The resulting curve illustrates the worsening of the fit $10 \lesssim \ell \lesssim 40$. {\color{black} Let us point out, however, that, within the framework discussed above \cite{Gaztanaga:2024vtr,Gazta_aga_2025}, tensor modes are not required to produce a similar sawtooth structure in the power spectrum at larger $\ell$.}
 
 \begin{figure}[ht]

\centering
\includegraphics[width=9.5cm]{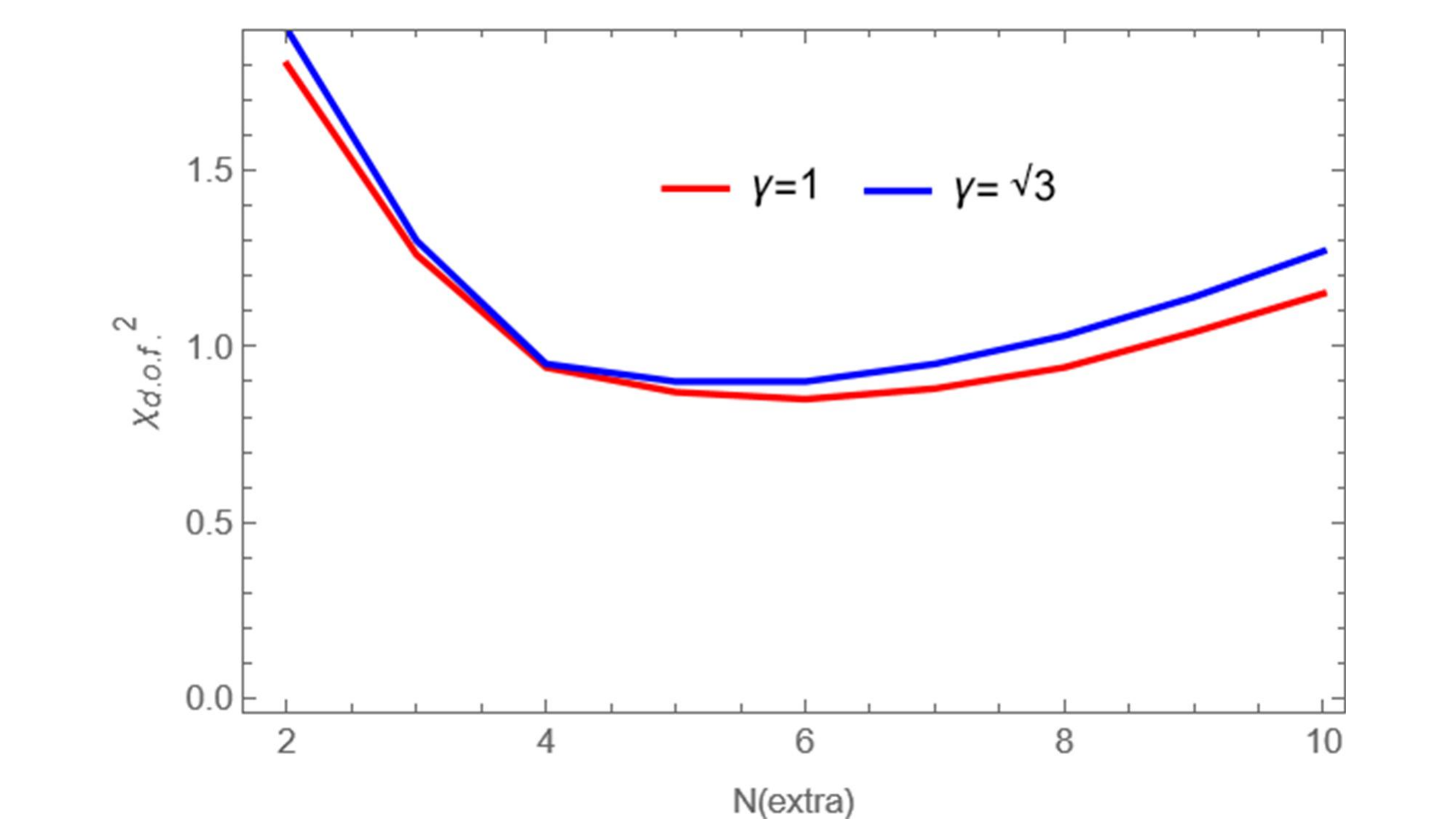}
\caption{\label{fig:8} $\chi^2_{\rm d.o.f.}$ obtained from different curves of $Q^{\rm TT}(\ell_{\rm max})$, as a function of the number of compactified EDs. A minimum is found at $N_{\rm extra}=6$ for both $\gamma=1$ (blue) and $\gamma=\sqrt{3}$ (red) according to Table II. No error bars are shown; the curves are intended to be indicative only.}

\end{figure}

\begin{table}[h!]
\caption{$\chi^2_{\rm d.o.f.}$ from the plot corresponding to the $Q^{\rm TT}$ statistic shown in Fig.~\ref{fig:6}, versus the possible number of tori and compactified extra dimensions (in brackets), as assumed in this work. Left/Right values correspond to $\gamma=1$ and $\gamma=\sqrt{3}$, respectively.\\}\label{tab:ToriDTT}
\begin{tabular}{|l|c|c|c|c|c|}

\hline 

\textbf{\# tori(ED)}: & \textbf{1(2)} & \textbf{2(4)} & \textbf{3(6)}&  \textbf{4(8)} & \textbf{5(10)}\\ 

\hline 

$\chi^2_{\rm d.o.f.}$ &  1.8/1.9 & 0.94/0.95 &  0.85/0.90 & 0.94/1.03 & 1.15/1.27 \\

\hline

\end{tabular}

\end{table}

\begin{table}[h!]
{\color{black}
\caption{$\chi^2_{\rm d.o.f.}$ from the plot corresponding to the $R^{\rm TT}$ statistic shown in Fig.~\ref{RTT-2}, versus the possible number of tori and compactified extra dimensions (in brackets), as assumed in this work. Left/Right values correspond to $\gamma=1$ and $\gamma=\sqrt{3}$, respectively.\\}\label{tab:toriRTT}
\begin{tabular}{|l|c|c|c|c|c|}

\hline 

\textbf{\# tori(ED)}: & \textbf{1(2)} & \textbf{2(4)} & \textbf{3(6)}&  \textbf{4(8)} & \textbf{5(10)}\\ 

\hline 

$\chi^2_{\rm d.o.f.}$ &  1.3/1.9 & 0.82/0.92 &  0.81/0.92 & 0.82/0.93 & 0.83/0.95 \\

\hline

\end{tabular}}

\end{table}

 \section{Testing the number of extra dimensions }\label{sec:Nextra}

 In the analysis below, we allow for the possibility that the number of tori ($N_{\rm tori}$) differs from three, and therefore that the number of extra dimensions ($N_{\rm extra}$) differs from six. As will be shown, a detailed parity analysis of the low-$\ell$ multipoles indicates a preference for $N_{\rm tori} = 3$, or equivalently $N_{\rm extra} = 6$, within our theoretical framework; this emerges as a suggestive result of the present work.

In Table \ref{tab:ToriDTT}, the values of $\chi^2_{\rm d.o.f.}$ obtained under different theoretical assumptions are presented as a function of the number of tori, $N_{\rm tori}$, based on the results shown in Fig.~\ref{fig:6}. Two values of the parameter $\gamma$, defined in Eq.~\eqref{ratioR}, have been considered, leading to the same conclusion: the smallest $\chi^2_{\rm d.o.f.}$ value (interpreted as indicating the best goodness of fit) is obtained in both cases for $N_{\rm tori}=3$.  In Fig.~\ref{fig:8}, we plot $\chi^2_{\rm d.o.f.}$ as a function of the number of compactified extra dimensions. The curves are intended to be illustrative, exhibiting a convex behavior with a minimum at $N_{\rm extra}=6$; no error bars are shown. \color{black} In Table \ref{tab:toriRTT} we show an equivalent study for the parity statistic $R^{\rm TT}$, leading to similar conclusions. \color{black}

{\color{black} A notable agreement is found between the results presented in the above tables and plots and the expectations of string-motivated theories, which posit a total of ten spacetime dimensions, nine of which are spatial, with six compactified dimensions. Nonetheless, these results should be interpreted with great caution, as our analysis relies on a simplified model involving several assumptions, in addition to the large observational uncertainties. In fact, the results only allow us to draw conclusions about the relative flatness of the curves shown in the two tables and should therefore be regarded as purely indicative of a preference for a number of extra dimensions close to six.}

\begin{figure}[ht]

\centering
\includegraphics[width=8.8cm]{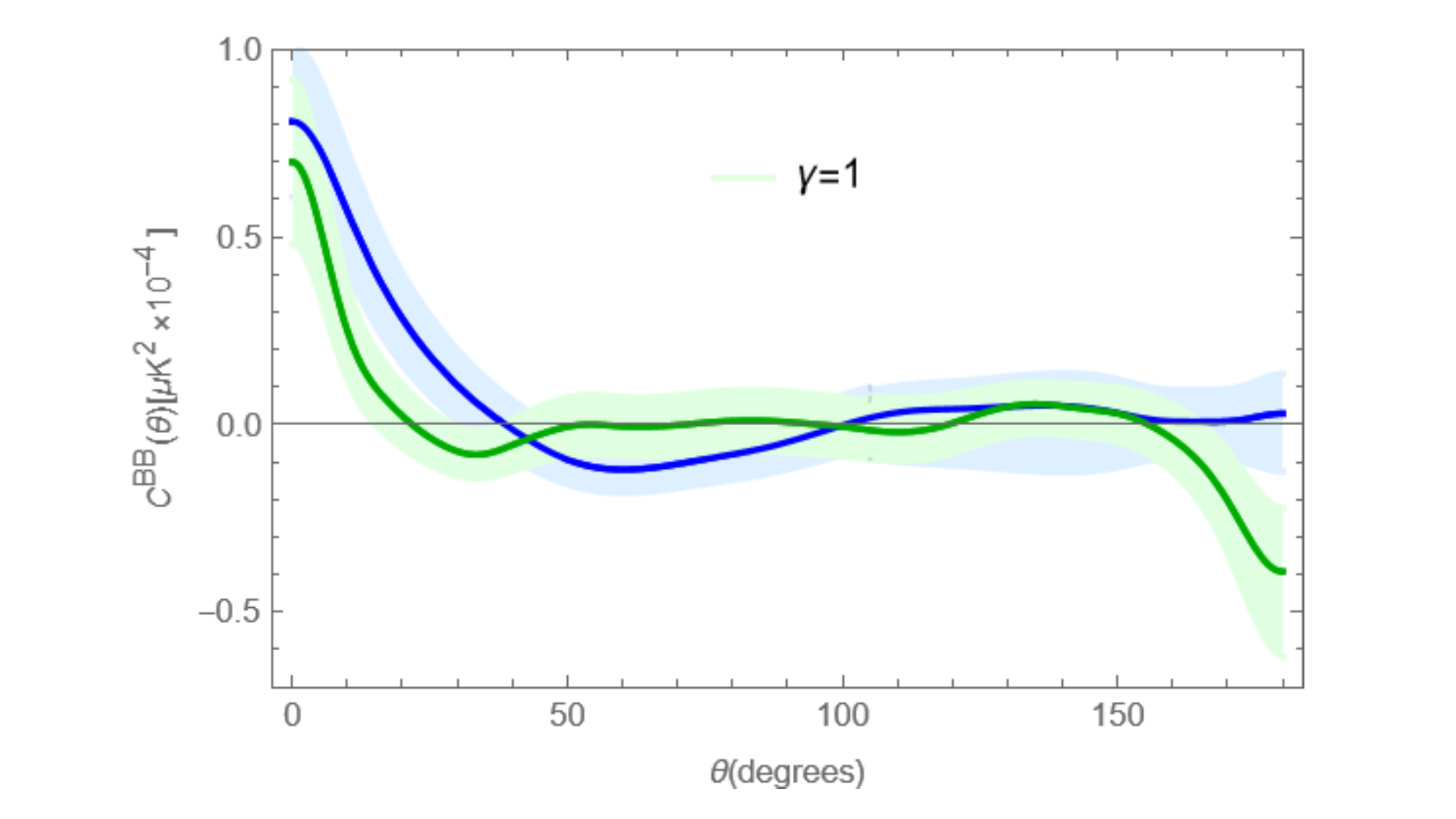}
\includegraphics[width=8.8cm]{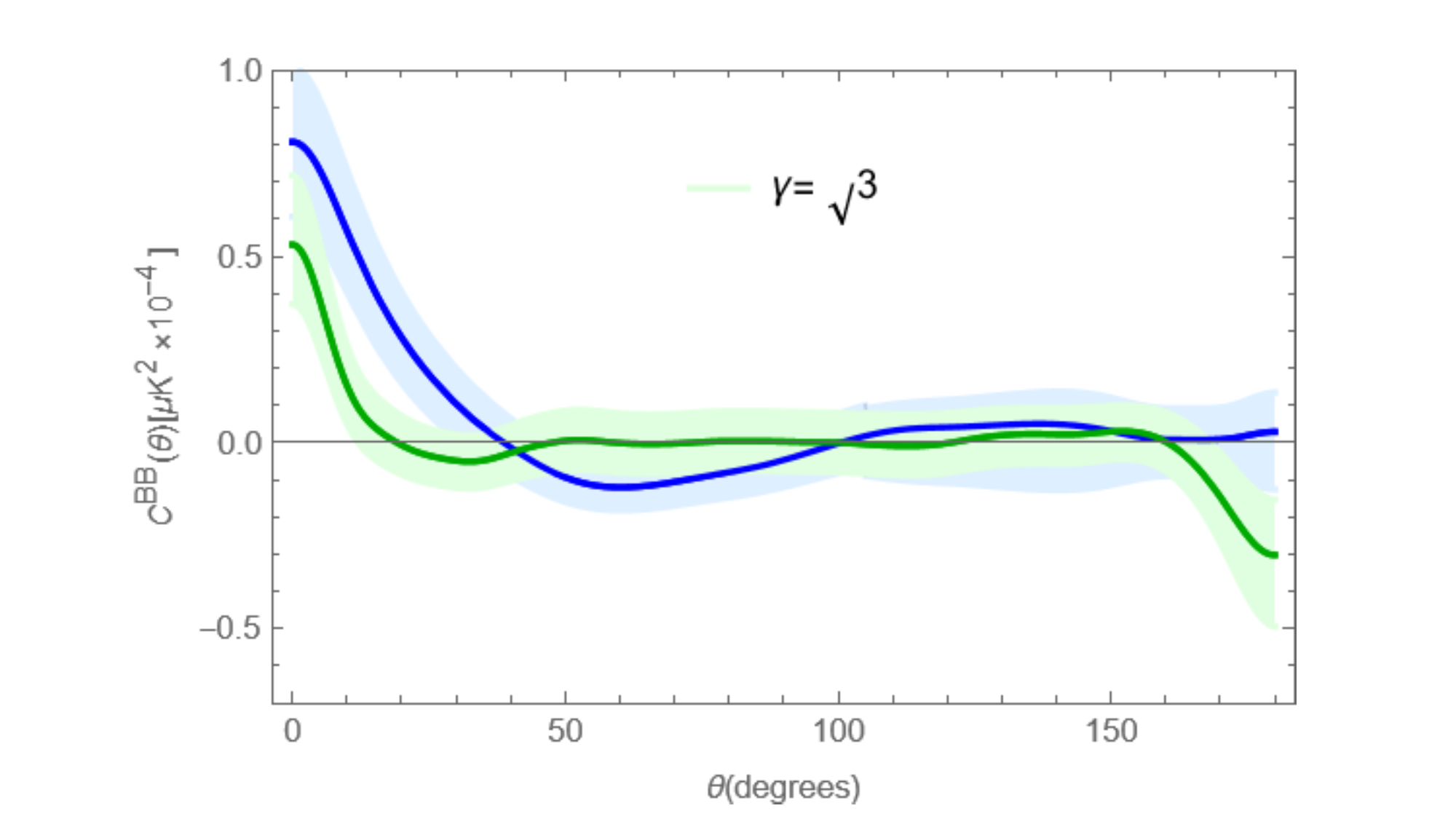}
\caption{\label{fig:9} B-mode polarization two-point auto-correlation function $C^{\rm BB}(\theta)$ (in units of $\mu{\rm K}^2 \times 10^{-4}$) for $\ell < 30$. Left panel: $\gamma = 1$; right panel: $\gamma = \sqrt{3}$. The blue curve corresponds to the case without cutoffs, while the green curve includes tensor IR cutoffs obtained from previous fits to the temperature power spectrum. The shaded bands represent cosmic variance, assuming ideal performance without instrumental noise or gravitational lensing. A distinctive signature emerges when comparing small and large angular scales.}
\end{figure}

\section{B-mode polarization prospective study}\label{sec:bmode} 

So far, our analysis has focused on the temperature anisotropies of the CMB, including the contribution of tensor modes from different EDs within a toroidal compactification of the early Universe. Particular emphasis was placed on the parity analysis at low-$\ell$ multipoles, where such subdominant effects are expected to be more clearly visible.
In what follows, we turn our attention to the B-mode polarization of the CMB as an alternative and independent test of our working hypotheses, namely, the influence of early Universe topology on angular correlations. 

As is well known, primordial B-mode polarization of the CMB can be generated at linear order only by tensor perturbations. However, B modes can also arise indirectly through gravitational lensing, which converts a fraction of the primordial E-mode polarization into B modes. In an ideal scenario, once the lensing contribution has been accurately removed from the data, the detection of primordial B-mode polarization would constitute a smoking gun for the existence of primordial gravitational waves and, potentially, for inflation itself. In contrast, scalar (density) perturbations generate only E-mode polarization at linear order in cosmological perturbation theory and therefore do not produce B modes at that order in the CMB.

On the other hand, primordial tensor modes predominantly generate B-mode polarization at large angular scales, contributing mainly to the low-$\ell$ region of the CMB power spectrum ($\ell \lesssim 100$). This signal includes both the reionization bump at very low multipoles, and the recombination bump around $\ell \sim 80$. In contrast, gravitational lensing of E-mode polarization by intervening large-scale structure induces secondary B modes that dominate at intermediate and high multipoles ($\ell \gtrsim 100$). Consequently, the observed B-mode spectrum consists of a superposition of the primordial tensor contribution at low $\ell$ and lensing-induced B modes at higher $\ell$, and a precise separation of these components is essential for inferring the presence of primordial gravitational waves.

Although B-mode polarization is intrinsically a spin-2 field \cite{PhysRevD.55.1830,Kamionkowski:2015yta}, we directly represent the BB angular auto-correlation function using a Legendre expansion (rather than spin-2 harmonics), as in the first part of this work. This approach serves as a convenient shortcut to highlight the parity-breaking effects observed in temperature correlations.

Moreover, this approximation can be justified by noting that one can construct rotational invariant combinations of the Stokes parameters (characterizing the linear polarization of the radiation field) whose correlation functions behave effectively as scalars \cite{Achucarro:2022qrl}. While the full and rigorous treatment requires the use of the CAMB package \cite{Lewis_2000}, the essential physical information is encoded in the multipole coefficients, which capture the impact of features such as infrared cutoffs and tensor-mode contributions arising from different compactified extra dimensions.

As in our previous analysis of temperature angular correlations, we again focus on large angular scales in the study of B-mode polarization, using the two-point correlation function $C^{\rm BB}(\theta)$ defined as: \cite{Baumann:2009mq,Liu:2024mvp,Liu:2025yvp}
\begin{equation}\label{eq:CBB}
C^{\rm BB}(\theta) = \sum_{\ell \ge 2} \frac{(2\ell+1)}{4\pi} \frac{(\ell+2)!}{(\ell-2)!}\  C_{\ell}^{\rm BB}\  P_{\ell}(\cos {\theta})\;,
\end{equation}
expanded in terms of Legendre polynomials.  
 We have restricted our study to low multipoles, i.e. $\ell < 30$, as a potential tool to distinguish between different possible topologies of the early Universe.

 In Fig.~\ref{fig:9}, we present two representative plots of $C^{\rm BB}(\theta)$ for $\gamma = 1$ (left panel) and $\gamma = \sqrt{3}$ (right panel), obtained using the tensor IR cutoff doublets inferred from previous fits to the temperature power spectrum. A tensor-to-scalar ratio of $r = 0.02$ was adopted. The curves can be straightforwardly rescaled for different values of $r$ to obtain approximate results.

A parity asymmetry between large and small angular scales is clearly visible in both plots, arising from the dominance of odd multipoles already observed in temperature correlations. The distinct behavior of the $C^{\rm BB}(\theta)$ curve across angular scales provides a characteristic signature of our prediction. Similar results and conclusions were reported in \cite{Sanchis-Lozano:2025csn} for the case of a single extra dimension; here, we extend the analysis to six compactified dimensions.

We stress that although the underlying physics is common to both temperature and B-mode polarization analyses, the associated phenomenology and detection methodologies differ. Consequently, the observation of a similar asymmetry in polarization would provide significant support for the hypothesis advanced in this work regarding extra dimensions. 

Admittedly, there is currently no robust observational evidence for B-mode polarization, so these theoretical predictions cannot yet be firmly confirmed. Nonetheless, forthcoming high-precision measurements of the CMB temperature and polarization both ground-based and space-borne are expected to shed light on this tantalizing possibility. 

In the former case, we first cite {\it LiteBIRD}, a space mission dedicated to full-sky CMB polarization measurements, aiming to detect large-scale B-modes from primordial gravitational waves with very high sensitivity \cite{LiteBIRD:2022cnt}. We also consider PICO \cite{hanany2019picoprobeinflationcosmic}, a proposed next-generation CMB polarization satellite designed to measure E- and B-modes over the full sky with ultra-high sensitivity, capable of probing primordial B-modes down to tensor-to-scalar ratios $r \sim 10^{-4}$ tightly constraining inflationary physics and foregrounds.

 On the other hand, the BICEP/Keck Array \cite{keckcollaboration2024} is a ground-based experiment targeting degree-scale B-mode polarization, paving the way for future, more sensitive measurements that probe primordial gravitational waves and refine constraints on inflation. We also mention balloon-borne missions such as SPIDER \cite{spidercollaboration2021constraintprimordialbmodesflight} and EBEX \cite{EBEX:2017oys}, which target CMB polarization from the stratosphere, reducing atmospheric noise and improving sensitivity to primordial B-modes.

\color{black}

\section{Discussion and comparison with alternative approaches}

Let us recapitulate some of the main observational facts and theoretical considerations underlying our work, together with possible alternative interpretations. We attribute the observational preference for odd-parity multipoles in the Legendre expansion of CMB angular correlations to the effects of extra dimensions on our four-dimensional world. Within our framework, the influence of scalar modes on the power spectrum is largely confined to the quadrupole and octupole contributions. Consequently, tensor modes are required in our approach to affect higher, although still low, multipoles. This feature follows directly from the cutoff structure adopted for the scalar sector and should not be interpreted as a general statement about the ability of scalar perturbations to generate parity asymmetry.

It is also worth emphasizing that parity asymmetry is not a new observational feature. An excess of power in odd over even multipoles at large angular scales has been reported in several analyses of WMAP and Planck temperature data [e.g. \cite{Hansen:2011wk,Kim2012,Aluri:2017mhj}]. These studies provide useful observational context for the present analysis. We stress again that such parity asymmetry does not, by itself, contradict the statistical isotropy of the Universe, since an alternating pattern in the angular power spectrum can arise without introducing a preferred spatial direction.

On the other hand, a totally different framework initially proposed in \cite{Kumar:2023lwj} leads to a qualitatively similar parity-asymmetric power spectrum and, consequently, to related features in the temperature correlations. The corresponding analysis in \cite{Gaztanaga:2024vtr} also yields a sawtooth-like structure in the temperature power spectrum, with an enhancement of odd relative to even multipoles. Thus, the qualitative similarity between the low-$\ell$ parity patterns found in the two approaches is noteworthy, despite their different theoretical frameworks. In particular, both exhibit an alternating enhancement and suppression of neighboring odd and even multipoles, although the physical origins of this pattern differ. In our scenario, it arises from the different infrared cutoffs associated with the BCs in the compact dimensions, whereas in Ref.~\cite{Gaztanaga:2024vtr} it originates from a two-component direct-sum inflationary framework that produces an odd-parity excess. Subsequent work on unitary quantum gravitational physics further develops this scenario \cite{Gazta_aga_2025,Gaztanaga:2025awe}.

Regarding CMB B-mode polarization, both approaches would lead to similar observational signatures. Nevertheless, since the underlying physical mechanisms differ, they could also give rise to distinct observable consequences. In the absence of sufficiently precise data, we can only speculate at this stage that cross-correlations (not examined in the present paper) might help discriminate between the two scenarios. This possibility deserves further investigation in future work.

Finally, we note that our approach is designed to account for the overall odd--even imbalance and its scale dependence at large angular scales, but it does not predict the particular localized or directional correlations discussed in some studies \cite{Creswell:2021eqi}. In our scenario, the parity asymmetry arises from the BCs in the compact dimensions, while statistical isotropy in the observable three-dimensional space is preserved; hence, no preferred spatial direction is introduced. Such directional features, if confirmed, would therefore provide additional information that could help discriminate between different physical origins of the CMB parity asymmetry.\color{black}

\section{Conclusions}\label{sec:concl}

As argued in previous works \cite{sanchis-sanz:2024,Sanchis-Lozano:2025csn}, where CMB temperature and polarization angular correlations were analyzed, both scalar and tensor modes may provide valuable insights into the geometry and topology of the early Universe. In this context, we have proposed a possible connection between the lack of large-angle correlations, the observed odd-parity preference in the CMB, and the existence of Kaluza--Klein extra dimensions, which could leave imprints on the temperature and polarization anisotropies observed today.

Our present analysis is based on a string-motivated framework involving toroidal compactification, in which the six-dimensional internal manifold is taken to be the product of three two-dimensional tori. This setup gives rise to infrared cutoffs that distinguish between odd and even multipoles in the scalar and tensor contributions to the primordial CMB angular power spectra. The resulting odd--even asymmetry thus emerges naturally from the assumed boundary conditions and affects the CMB anisotropies at large angular scales.

We emphasize that, although implicitly assumed throughout this work, our starting theoretical framework does not strictly require an inflationary epoch: the orbifold compactification of the extra dimensions and the emergence of the associated IR cutoffs are assumed to occur after the Planck era and before the onset of inflation. Nonetheless, in the absence of inflation or a viable alternative mechanism, it remains challenging to account for the full set of observed properties of the CMB power spectrum, including its near scale invariance and acoustic peak structure.

Following our phenomenological approach, we first fitted the CMB angular power spectrum at $\ell < 30$ using Planck data, obtaining estimates of the scalar and tensor IR cutoffs under the assumption of a tensor-to-scalar ratio $r=0.02$. This analysis is closely related to fitting the temperature two-point correlation function, but focuses on the low-multipole regime.

We then performed a consistency check based on a goodness-of-fit analysis of two closely related parity statistics, which supports an odd-parity preference at low multipoles. Moreover, the results are compatible with a scenario involving six compactified extra dimensions, as expected in our string-motivated framework. We stress, however, that the present data do not warrant a discovery claim; rather, they may point to a possible imprint of the topology of the early Universe that merits further investigation with forthcoming observations.

Parity-violating new physics can also be investigated within other theoretical frameworks and through different observables \cite{Lue:1998mq,Feng:2004mq,Liu:2006uh,Saito:2007kt,Contaldi:2008yz}. More recently, Refs.~\cite{Gazta_aga_2025,Gaztanaga:2025awe} have obtained a scalar power-spectrum pattern qualitatively similar to ours, despite the substantially different underlying physical mechanisms. \color{black}Therefore, we point out that the observed CMB parity asymmetries in temperature and polarization, so far by themselves, do not constitute unique evidence for extra dimensions, and further investigation is required to establish their physical origin.\color{black}

Following the same strategy, we have extended our analysis to the B-mode polarization auto-correlation function within the same framework used for the temperature anisotropies. At present, robust observational constraints on primordial B-mode polarization at low multipoles remain limited. We therefore present a prediction for the two-point $BB$ correlation function, which exhibits a pattern similar to that of the temperature anisotropies. In our framework, both arise from the same underlying mechanism: an odd--even asymmetry induced by the assumed toroidal compactification and its associated boundary conditions. If a consistent odd-parity preference were eventually established across temperature and polarization, including both $E$- and $B$-modes, this would provide an important cross-check of the proposed scenario, although its statistical significance and possible alternative explanations would need to be carefully assessed.

Forthcoming high-precision measurements of CMB temperature and polarization, from space missions such as {\it LiteBIRD} and PICO, balloon-borne experiments such as SPIDER and EBEX, and ground-based experiments such as BICEP/Keck, may provide new insights into the possibility of probing the topology of the early Universe and associated non-standard physics. In particular, improved measurements could constrain the IR cutoff parameters more tightly, helping to discriminate among competing models. A potential detection of primordial gravitational waves through their B-mode polarization signature would provide an additional and complementary constraint on our framework.

Finally, our analysis could be extended, albeit non-trivially, to other types of compactification in string theory, including brane-based scenarios and compactifications involving Calabi--Yau manifolds. Such extensions would require a detailed treatment of the relevant internal geometry, mode spectrum, and boundary conditions to establish whether analogous parity-dependent IR cutoffs arise. We hope to investigate these possibilities in future work.

\section*{Acknowledgments}

\color{black} We thank the anonymous referee for a thorough review of the manuscript and for insightful comments that have helped us improve the paper\color{black}.  N.E.M. would like to thank the University of Valencia and its Theoretical Physics Department for a visiting Research Professorship
supported by the programme  \emph{Atracci\'on de Talento}
INV25-01-15 and departmental funds, which made the collaboration leading to the present work possible. 
The work of NEM is supported in part by the UK Science and Technology Facilities research Council (STFC) under the research grant no. ST/X000753/1. 
 M.A.S.L. acknowledges  support by the Spanish Agencia Estatal de Investigacion, under Grant PID2023-151418NB-I00 funded by MCIU/AEI/10.13039/501100011033/ FEDER, UE, and by GV under grant CIPROM/2022/36.
NEM also acknowledges participation in the COST Association Actions CA21136 “Addressing observational
tensions in cosmology with systematics and fundamental physics (CosmoVerse)” and CA23130 ``Bridging high and low
energies in search of Quantum Gravity (BridgeQG)”.

\appendix

\section{Orbifold compactification: physical interpretation, notation and code}
\label{sec:appA}

{\color{black}

\subsection{Physical interpretation}

The emergence of IR cutoffs in the power spectrum, originating from extra dimensions but manifesting themselves in the four-dimensional effective theory, plays a central role in our work. We therefore summarize the underlying mechanism here.

First, we note that the presence of a lowest allowed momentum in a compact KK dimension modifies the spectrum of momenta accessible to the four-dimensional effective theory. To illustrate this point with a simple example, consider the propagator of a scalar field in four-dimensional spacetime. For simplicity, retaining only one momentum dimension, the corresponding momentum-space integral can be schematically written as
\begin{equation}
\int_0^\infty \frac{dk}{k^2+m^2}
\quad\longrightarrow\quad
\int_{k_{\min}}^\infty \frac{dk}{k^2+m^2},
\end{equation}
where $k_{\min}$ denotes the lowest allowed momentum and therefore acts as an infrared cutoff.

In the regime $k\gg m$, the integrand approaches the massless form, so that
\begin{equation}
\frac{1}{k^2+m^2}\simeq \frac{1}{k^2}.
\end{equation}
Consequently, the corresponding infrared contribution is regulated by the lower bound,
\begin{equation}
\int_{k_{\min}}^\infty \frac{dk}{k^2}.
\end{equation}
This simple example makes explicit that the effect of the compact extra dimension is to exclude modes with momenta below the lowest allowed value $k_{\min}$. In the effective four-dimensional description, this manifests itself as an infrared cutoff in the momentum spectrum.

In our specific construction, the value of $k_{\min}$ is not introduced as an independent phenomenological cutoff. Rather, it is inherited from the spectrum of the compact extra dimensions. Different combinations of boundary conditions on one or more extra dimensions lead to distinct KK towers, with different lowest allowed KK momenta and, equivalently, lowest KK masses. As developed in the main text, upon dimensional reduction these lowest spectral scales provide the corresponding IR scales in the four-dimensional effective description.

In particular, the BC combinations considered in our construction are associated with the odd- and even-parity sectors and therefore lead to different lowest KK scales for these two sectors. We consequently obtain different effective IR cutoffs, $k_{\min}^{\rm odd}$ and $k_{\min}^{\rm even}$, in the corresponding contributions to the primordial power spectrum. These cutoffs are then propagated into the multipole coefficients $C_\ell$ entering the CMB two-point angular correlation function. Thus, the parity dependence of the IR cutoff originates from the extra-dimensional boundary conditions and is subsequently manifested as a parity imbalance in the observed CMB angular correlations, as postulated in this and our previous works.

The effect is particularly pronounced at large angular separations, where the lowest multipoles provide the dominant contribution to the correlation function. In particular, the odd-parity dominance produces a characteristic behavior in the vicinity of the antipodal region, as explained and illustrated in Fig.~2 in the main text.}

\subsubsection*{\color{black}Extension to more extra dimensions}

On the other hand, the theoretical framework of this work is based on a string-inspired scenario with an orbifold compactification of three tori, labeled by $i = 1,2,3$. This setup gives rise to three pairs of extra dimensions which, together with the usual three spatial dimensions, yield a total of nine spatial dimensions, as expected in superstring theory. For completeness and comparison, we shall also consider the possibility that the number of tori differs from three, and consequently that the number of orbifold-compactified dimensions differs from six.

We characterize each compactified ED associated with the $i$-th torus by a radius $R_{i,I}$, where $I = 1,2$ labels the two orbifold-compactified dimensions within each torus. For each ED, we distinguish between scalar and tensor modes, as well as between even- and odd-parity modes. These length (or energy) scales define a hierarchical structure of IR cutoffs across the different combinations. In particular, a smaller radius corresponds to a larger IR cutoff in energy units.

Each IR cutoff is related to a dimensionless variable denoted by $u_{\rm min}$ through the expression
\begin{equation}
k_{\rm min}^{\rm even/odd}(\mathrm{scalar/tensor}) = \frac{u_{\rm min}^{\rm even/odd}({\rm scalar/tensor})}{r_L}\;,
\end{equation}
where $r_L$ denotes the comoving distance to the LSS at which the CMB was released. As usual, the finite time width of decoupling is neglected in our calculations. 

The numerical values of the dimensionless parameters for even and odd modes, $u_{\rm min}^{\rm even/odd}$, respectively, which enter as lower limits in the computation of the multipole coefficients, are obtained from fits to CMB angular correlation data, as explained in the main text.

\subsection{\color{black} Notation}
The following choices were adopted in our study:
\begin{itemize}
\item $q = 2$, following the boundary conditions imposed on each compact ED.
\item $\alpha = 1$ for a flat geometry, whereas $\alpha \gtrsim 2$ for a warped geometry.
\item $\beta = 2$, as preferred by our phenomenological study.
\item $\gamma = 1$ and $\gamma = \sqrt{3}$, the latter corresponding to a hexagonal torus.
\end{itemize}

We now extend this notation to include the additional EDs:
$u^{\rm even/odd}_{\rm min}(i,I)$, with $i = 1,2,3$ and $I = 1,2$. For notational clarity, we replace the subscript ${\rm min}$ (which will be understood) by $S$ or $T$ to denote scalar and tensor modes, respectively. Thus, we will hereafter write $u_{S/T}^{\rm even/odd}(i,I)$ to denote the dimensionless lower limits in the integrals entering the computation of the coefficients $C_{\ell}$ (see Eq.~\eqref{eq:Cellcutoffs}). 

We define the following ratios relating the IR cutoffs among different radii of the EDs:
\begin{equation}\label{eq:ratios}
    \frac{u_{S/T}^{\rm even}(i,I)}{u_{S/T}^{\rm odd}(i,I)}=q\ ;\ \ 
    \frac{u_{T}^{\rm even/odd}(i,I)}{u_{S}^{\rm even/odd}(i,I)}=\alpha\ ;\ \
    \frac{u_{T}^{\rm even/odd}(i+1,I)}{u_{T}^{\rm even/odd}(i,I)}= \beta\ ;\ \ \frac{u_{T}^{\rm even/odd}(i,2)}{u_{T}^{\rm even/odd}(i,1)}= \gamma\ ;\ \ \ i=1,2,3;\ I=1,2
\end{equation}

\subsection*{Code}\label{sec:code}

To perform our analysis of temperature and polarization angular correlations within the framework of toroidal compactification, we employed a numerical code to compute the scalar and tensor multipole coefficients entering the Legendre expansions in Eqs.~\eqref{eq:C2}, \eqref{eq:CEE}, and \eqref{eq:CBB}. Specific values of the parameters $q$, $\alpha$, $\beta$, and $\gamma$ were selected based on the best fit to the available observational data. These values were obtained through an optimization procedure using temperature anisotropy data, as described in the main text.

For concreteness we provide the numerical values for the lowest odd and even IR cutoffs for scalar and tensor modes:
\begin{eqnarray}
k_{\rm min}^{\rm odd/even}(\rm scalar) &=& \frac{u_{\rm min}^{\rm odd/even}(\rm scalar)}{r_L} = 1.45/2.90 \times 10^{-4}\ {\rm Mpc}^{-1} \\
k_{\rm min}^{\rm odd/even}(\rm tensor) &=& \frac{u_{\rm min}^{\rm odd/even}(\rm tensor)}{r_L} = 2.90/5.80 \times 10^{-4}\ {\rm Mpc}^{-1} 
\end{eqnarray}
which correspond to the largest radius of the set $\{R_{iI}\}$, i.e. $i=I=1$ according to the conventions of Section II, in particular \eqref{m2Rij}. The code also incorporates those higher IR cutoffs, corresponding to smaller radii, which have a sizeable impact on the power spectrum $\ell < 30$.

\begin{figure}[ht]

\centering
\includegraphics[width=12.5cm]{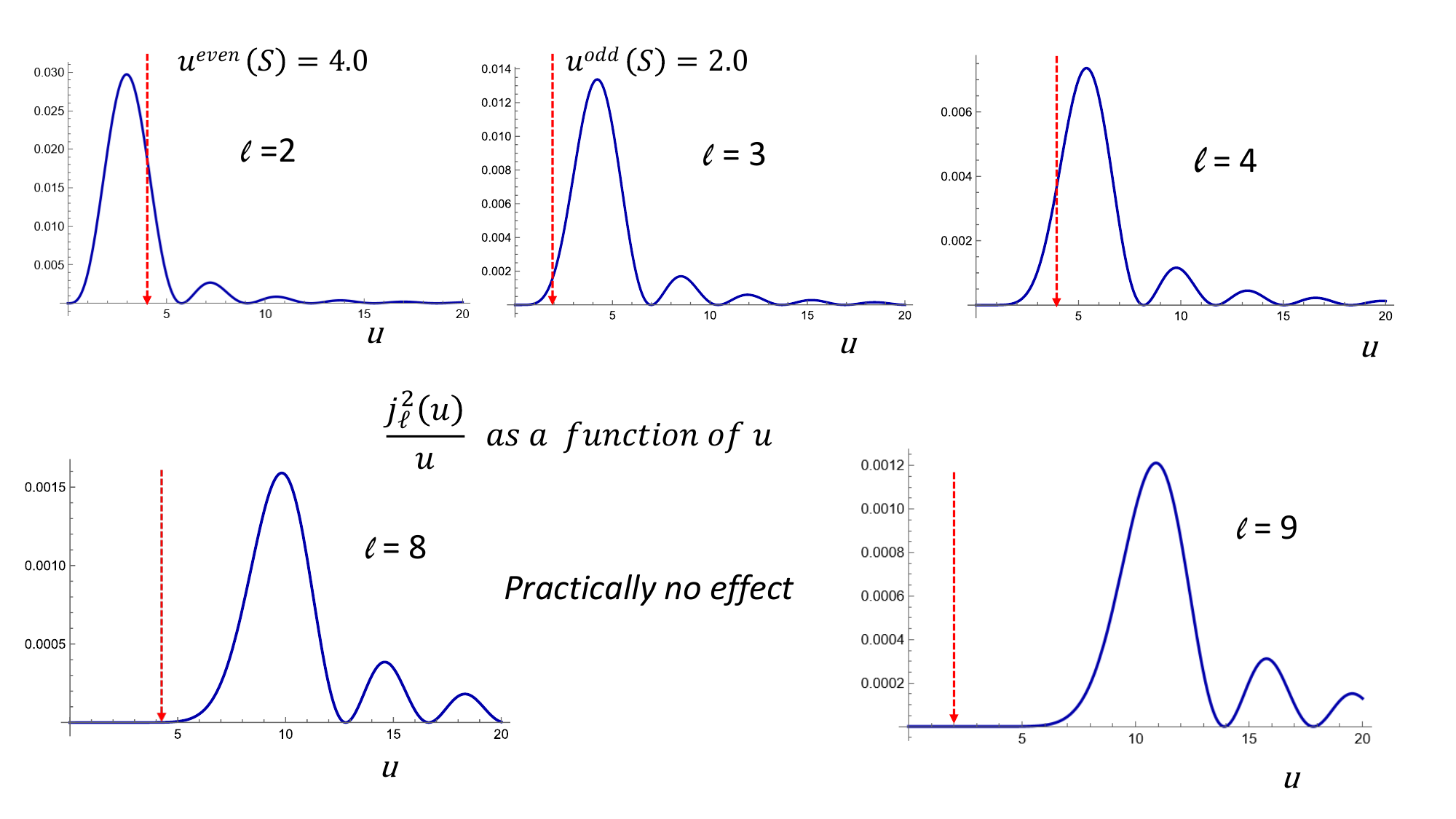}
\caption{\label{fig:10} Representative plots of the integrand of some low-$\ell$ multipole coefficients $C_\ell$ for scalar modes showing separately the effect of the sharp odd and even cutoffs. The effect becomes negligible for $\ell_{\rm odd/even} > u_{\rm min}^{\rm odd/even}$.}
\end{figure}

\begin{figure}[ht]

\centering
\includegraphics[width=12.5cm]{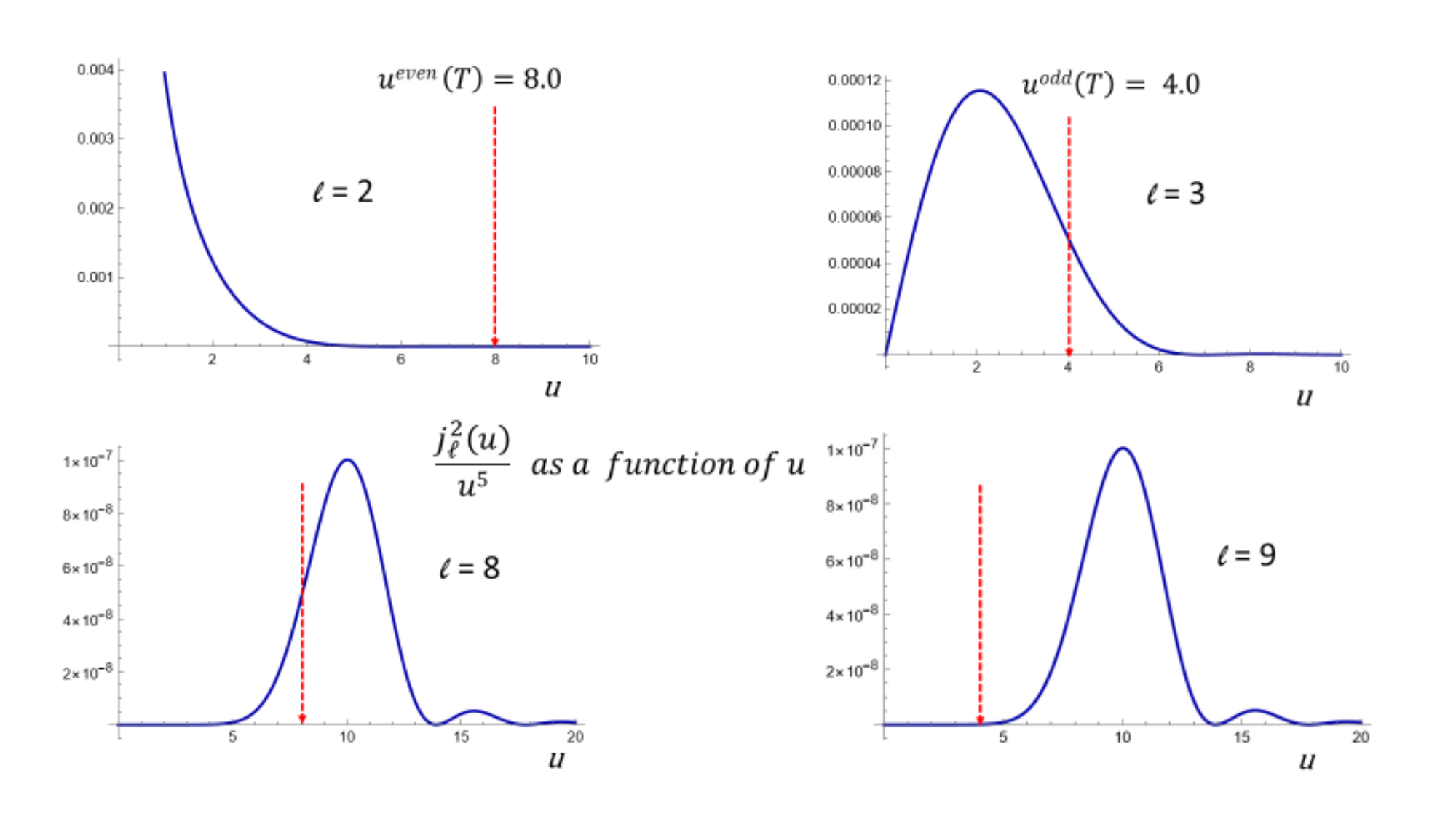}
\caption{\label{fig:11} The same as in Fig.\ref{fig:10} for tensor modes.}
\end{figure}

\section{Sharp versus smooth cutoff(s)}\label{sec:appB}

As commented in the main text, a sharp IR cutoff was introduced into primordial power spectrum in previous papers, thereby removing low frequencies directly affecting the computation of the corresponding multipole coefficients. Distinguishing odd and even scalar modes, we wrote in Eq.\eqref{eq:Cellcutoffs} and Eq.\eqref{eq:Celltensoroddeven}:
\begin{equation}
C_{\ell_{\rm odd/even}}^{\rm TT}{\rm (scalar)} = N_S \int_{u_{\rm min}^{\rm odd/even}\ {\rm (scalar)}}^{\infty} du\ \frac{j_{\ell}^2(u)}{u}\ ;\ \  C_{\ell_{\rm odd/even}}^{\rm TT}{\rm (tensor)} = N_T \int_{u_{\rm min}^{\rm odd/even}\ {\rm (tensor)}}^{\infty} du\ \frac{j_{\ell}^2(u)}{u^5}
\end{equation}
for scalar and tensor modes, respectively.

 The spherical Bessel function $j_\ell(u)$ is sharply peaked around $u \sim \ell$ \cite{Abramowitz:1964}, which underlies the usual correspondence $kr \sim \ell$ in angular projections. When dividing by powers of $u$, this peak is somehow shifted toward smaller values: for $j_\ell(u)/u$ the shift is mild, and the maximum still lies very close to $u \approx \ell$, whereas for higher powers, like $j_\ell(u)/u^5$, the peak moves more noticeably to $u < \ell$.

In any event, the main conclusion remains valid for both scalar and tensor modes: the net effect of a sharp lower cutoff on the coefficient $C_{\ell}$ is to slightly reduce its numerical value for multipoles satisfying $\ell_{\rm odd/even} \lesssim u_{\rm min}^{\rm odd/even}$, as shown in Fig.~\ref{fig:10}. Conversely, for $\ell_{\rm odd/even} > u_{\rm min}^{\rm odd/even}$, the effect becomes negligible.

A similar behavior is observed for tensor modes and their corresponding IR cutoffs, as illustrated in Fig.~\ref{fig:11}. However, due to the damping factor $1/u^5$, the suppression for $\ell \lesssim u_{\rm min}$ is more pronounced for tensor than for scalar modes. This feature becomes particularly relevant when constructing the angular power spectrum.

On the other hand, we have incorporated a smoothing of the hard lower limits by means of the function
{\color{black}
\begin{equation}\label{eq:fsmooth}
f(u; u_{\rm min}, \Delta_u)\ \frac{1-e^{-u/\Delta_u}}{1+e^{-(u-u_{min})/\Delta_u}} 
\; ,
\end{equation}}
where $\Delta_u$ denotes the width associated with the otherwise sharp cutoff at $u_{\rm min}$, for either odd/even and scalar/tensor modes. {\color{black} This definition of the smoothing function avoids the divergence of the integrand at $u=0$ for $\ell=2$ \footnote{{\color{black} We thank the anonymous referee for pointing out this issue with our previous choice of smoothing function. We note, however, that the numerical values of the corresponding integrals are practically identical in both cases.}}.

Despite the arbitrariness in the choice of the smoothing function, the above definition exhibits the expected behaviour, yielding results that are very similar to those obtained with a hard cutoff, both for the multipole coefficients and for the final results.}

Hence,
\begin{equation}\label{eq:Cellcutoffsm}
C_{\ell_{\rm odd/even}}^{\rm TT}{\rm (scalar)} = N_S \int_0^\infty du \ f(u; u_{\rm min}^{\rm odd/even}\ ,\Delta_u)\ \frac{j_{\ell}^2(u)}{u}\ \;\ .
\end{equation}
and similarly for tensor modes.

Throughout this work $\Delta_u$ was set equal to 0.2,  following the uncertainty in Eq.~\eqref{eq:ustnew}, for scalar and tensor modes. This choice can be regarded as a phenomenological parametrization of a finite-width transition in momentum space. It is qualitatively similar in sharpness to the turnover exhibited in pre-inflationary spectra derived in \cite{Ashtekar:2021izi}; however, in that case the detailed shape of the suppression follows from the underlying dynamics rather than from an imposed functional form.

The net effect of applying the smoothing function given in Eq.~\eqref{eq:fsmooth} to even/odd and scalar/tensor modes is to mitigate the impact of the hard cutoff on the computation of the multipole coefficients. Allowing $\Delta_u$ to vary over a reasonable range ($0.1$–$0.5$) has no significant impact on our results, as the hard cutoff can be adjusted accordingly.

\bibliographystyle{apsrev4-2}
\bibliography{Topologyrefs.bib}

\end{document}